\documentclass[journal]{IEEEtran}
\usepackage{times}
\usepackage{epsfig}
\usepackage{graphicx}
\usepackage{amsmath}
\usepackage{amssymb}
\usepackage{subfigure}
\usepackage{colortbl}
\usepackage{multirow}
\usepackage{soul}
\usepackage{adjustbox}
\usepackage{array}
\usepackage{setspace}
\ifCLASSINFOpdf
\else
\fi
\begin{document}
%
\title{BioTouchPass: Handwritten Passwords \\ for Touchscreen Biometrics}
%
%
%

\author{Ruben Tolosana,
        Ruben Vera-Rodriguez,~\IEEEmembership{Member,~IEEE,}
        and Julian Fierrez,~\IEEEmembership{Member,~IEEE}\\

\thanks{The authors are with the Biometrics and Data Pattern Analytics - BiDA Lab, Escuela Politecnica Superior, Universidad Autonoma de Madrid, 28049 Madrid, Spain (e-mail: ruben.tolosana@uam.es; ruben.vera@uam.es; julian.fierrez@uam.es).}}

%
%

\markboth{Journal of \LaTeX\ Class Files,~Vol.~13, No.~9, March~2016}%
{Shell \MakeLowercase{\textit{et al.}}: Bare Demo of IEEEtran.cls for Journals}
%



\maketitle

\begin{abstract}
This work enhances traditional authentication systems based on Personal Identification Numbers (PIN) and One-Time Passwords (OTP) through the incorporation of biometric information as a second level of user authentication. In our proposed approach, users draw each digit of the password on the touchscreen of the device instead of typing them as usual. A complete analysis of our proposed biometric system is carried out regarding the discriminative power of each handwritten digit and the robustness when increasing the length of the password and the number of enrolment samples. The new e-BioDigit database, which comprises on-line handwritten digits from 0 to 9, has been acquired using the finger as input on a mobile device. This database is used in the experiments reported in this work and it is available together with benchmark results in GitHub\footnote{https://github.com/BiDAlab/eBioDigitDB}. Finally, we discuss specific details for the deployment of our proposed approach on current PIN and OTP systems, achieving results with Equal Error Rates (EERs) ca. 4.0\% when the attacker knows the password. These results encourage the deployment of our proposed approach in comparison to traditional PIN and OTP systems where the attack would have 100\% success rate under the same impostor scenario. 
\end{abstract}


%
\IEEEpeerreviewmaketitle

\section{Introduction}
%
%
%
%

\IEEEPARstart{M}{obile} devices have become an indispensable tool for most people nowadays~\cite{MobileAddictive}. The rapid and continuous deployment of mobile devices around the world has been motivated not only by the high technological evolution and new features incorporated but also to the new internet infrastructures like 5G that allows the communication and use of social media in real time, among many other factors. In this way, both public and private sectors are aware of the importance of mobile devices for the society and are trying to deploy their services through user-friendly mobile applications ensuring data protection and high security. 

Traditionally, the two most prevalent user authentication approaches have been Personal Identification Numbers (PIN) and One-Time Passwords (OTP). While PIN-based authentication systems require users to memorize their personal passwords, OTP-based systems avoid users to memorize them as the security system is in charge of selecting and providing to the user a different password each time is required, e.g., sending messages to personal mobile devices or special tokens. Despite the high popularity and deployment of PIN- and OTP-based authentication systems in real scenarios, many studies have highlighted the weaknesses of these approaches~\cite{Bonneau_2012, Galbally_password_TIFS_2017}. First, it is common to use passwords based on sequential digits, personal information such as birth dates, or simply words such as ``password" or ``qwerty" that are very easy to guess. Second, passwords that are typed on mobile devices such as tablets or smartphones are susceptible to  ``smudge attacks", i.e., the deposition of finger grease traces on the touchscreen can be used for the impostors to guess the password~\cite{Aviv_2010}. Finally, password-based authentication is also vulnerable to ``shoulder surfing". This type of attack is produced when the impostor can observe directly or use external recording devices to collect the user information. This attack has attracted the attention of many researchers in recent years due to the increased deployment of handheld recording devices and public surveillance infrastructures~\cite{Shukla_2014, Yue_2014}. Biometric recognition schemes are able to cope with these challenges by combining both a high level of security and convenience~\cite{Surveying_mobile}.

\begin{figure*}[tb]
  \centering
    \includegraphics[width=\linewidth]{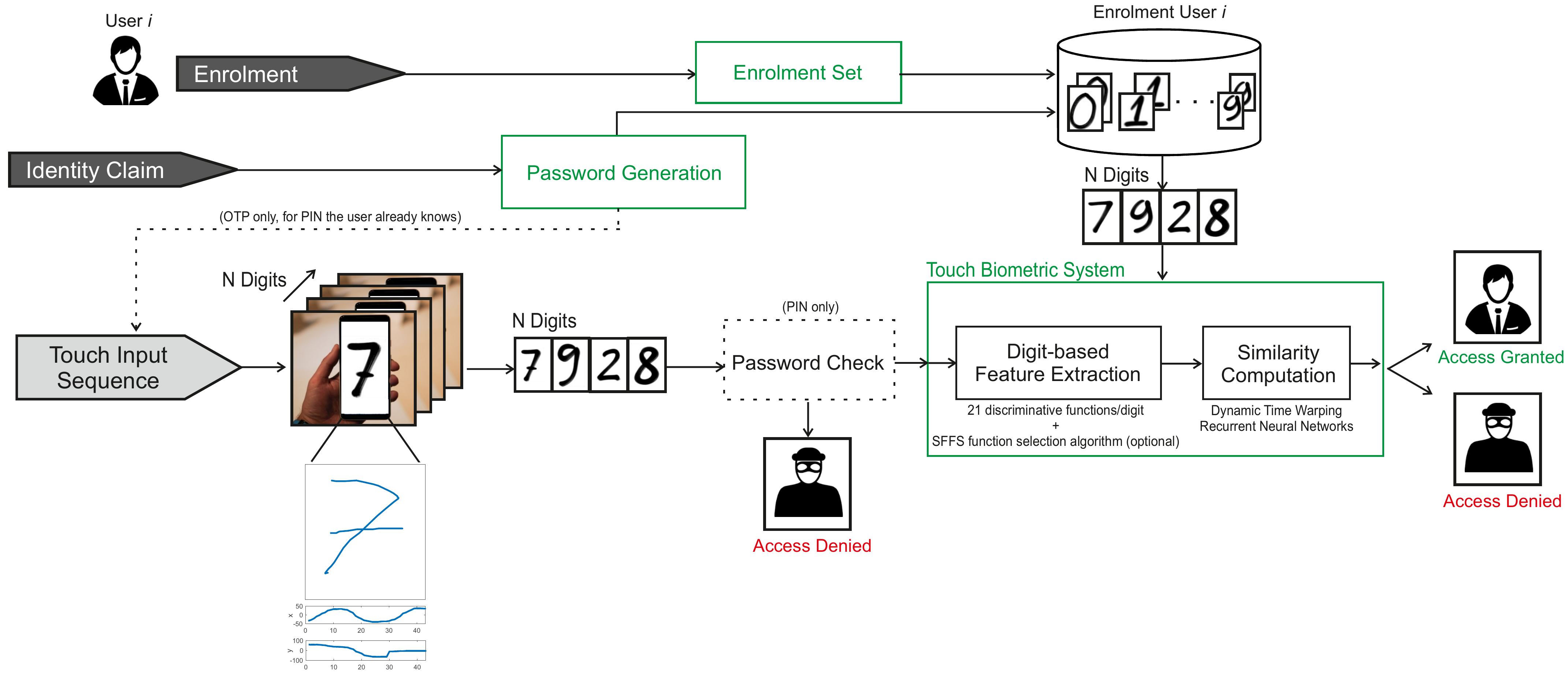}
  \caption{Architecture of our proposed password-based mobile authentication approach including handwritten touch biometrics in a two-factor authentication scheme applicable both to user-generated PIN and OTP systems.}
  \label{diagram_main}
\end{figure*}

This study evaluates the advantages and potential of incorporating biometrics to password-based mobile authentication systems, asking the users to draw each digit of the password on the touchscreen instead of typing them as usual. This way, the traditional authentication systems are enhanced by incorporating dynamic handwritten biometric information. One example of use that motivates our proposed approach is on internet payments with credit cards. Banks usually send a numerical password (typically between 6 and 8 digits) to the user's mobile device. This numerical password must be inserted by the user in the security platform in order to complete the payment. Our proposed approach enhances such scenario by including a second authentication factor based on the user biometric information while drawing the digits. Fig.~\ref{diagram_main} shows a general architecture of our proposed password-based mobile authentication approach. The three following main modules are analyzed in this study: \textit{i)} enrolment set, \textit{ii)} password generation, and \textit{iii)} touch biometric system. Depending on the final application (i.e., PIN or OTP), the handwritten digits can be first recognized using for example an Optical Character Recognition (OCR) system in order to verify the authenticity of the password. After this first authentication stage, the biometric information of the handwritten digits is compared in a second authentication stage to the enrolment data of the claimed user, comparing each digit one by one. In this study we focus on the second authentication stage based on the behavioral information of the user while performing the handwritten digits as the recognition of numerical digits has already shown to be an almost solved problem with errors close to 0\%~\cite{MNIST_2013, MNIST_2015}. Therefore, in this study we make the assumption that impostors pass the first stage of the security system (i.e., they know the password of the user to attack) and thus, the attack would have 100\% success rate if our proposed approach was not present.

The main contributions of this study are related to our proposed architecture, the competitive results obtained with respect to related research, and our experimental findings: 
\begin{itemize}
\item We survey and compare advantages and limitations of recent research on touch biometrics for mobile authentication.
\item We incorporate biometrics to password-based mobile authentication. Two different state-of-the-art approaches are studied for the similarity computation: \textit{i)} Dynamic Time Warping (DTW), which is widely used in many different fields such as handwriting and speech; and \textit{ii)} Recurrent Neural Networks (RNNs), which are specific deep learning architectures considered for modelling sequential data with arbitrary length. 
\item We perform a complete analysis of the touch biometric system regarding the discriminative power of each handwritten digit. In addition, we analyze the robustness of our proposed approach when increasing the length of the handwritten password and the number of available enrolment samples per user. 
\item We discuss specific details for the deployment of our proposed approach on current PIN- and OTP-based authentication systems, including various strategies for password generation.
\item We achieve better results than other verification schemes such as the handwritten signature and graphical passwords, as well as other recent works on touch biometrics.
\item We introduce the new e-BioDigit database, which comprises on-line handwritten numerical digits from 0 to 9 for a total of 93 users, captured on a mobile device using finger touch interactions. Handwritten digits were acquired in two different sessions in order to capture the intra-user variability. This database is publicly available to the research community.
\end{itemize}

The remainder of the paper is organised as follows. Sec.~\ref{related_works} summarizes related works in touch biometrics for mobile scenarios. Sec.~\ref{touch_biometric_system} describes our proposed touch biometric system. Sec.~\ref{e_BioDigit_description} describes the new e-BioDigit database, which comprises on-line handwritten numerical digits from 0 to 9. Sec.~\ref{experimental_protocol} and \ref{experimental_results} describe the experimental protocol and results achieved using our proposed approach, respectively. Sec.~\ref{password_real_systems} discusses specific details for the deployment of our proposed approach on current PIN- and OTP-based authentication systems, including password generation strategies. Finally, Sec.~\ref{conclusions} draws the final conclusions and points out some future work lines.

\begin{table*}[tb]
\centering
\caption{Comparison of different touch biometric approaches for mobile scenarios. Acc = Accuracy.}
\label{comparative_mobileScenarios}
\begin{adjustbox}{width=\textwidth}
\begin{tabular}{ccccccc}
\hline
\textbf{Study}                       & \textbf{Method}  & \textbf{Features} & \textbf{Classifiers}      & \multicolumn{2}{c}{\textbf{Verification Performance}} & \textbf{\# Participants} \\ 
                           &                &   &             & \textbf{Random Attack} & \textbf{Imitation Attack}                                                      & \textbf{(Dataset)}             \\ \hline \hline
Angulo \textit{et al.} (2011) \\ \cite{Angulo_2011}              & Lock Pattern Dynamics & Timing-related Features & Random Forest       & -     & EER = 10.39\%                                                                       & 32           \\ \hline
Lacharme \textit{et al.} (2016) \\ \cite{Lacharme_2016}              & Lock Pattern Dynamics & Dynamic Features & Hamming Distance       & -     & EER = 15.0\%                                                                       & 34           \\ \hline 
Zezschwitz \textit{et al.} (2016) \\ \cite{deLuca_lockPatterns}              & Lock Pattern Dynamics & Shape Features & Greedy Clustering       & -     & -                                                                      & 506           \\ \hline \hline
Buschek \textit{et al.} (2015) \\ \cite{deLuca_typing_2}              & Keystroke & Font Adaptation Features & Manual        & Acc = 94.8\%     & -                                                                      & 91          \\ \hline
Buschek \textit{et al.} (2015) \\ \cite{deLuca_typing_1}              & Keystroke & Touch-specific Features & GM, $k$NN, LSAD        & EER = 13.74\%     & -                                                                      & 28          \\ \hline \hline
Li \textit{et al.} (2013) \\ \cite{Li_2013}              & Touchscreen Gestures & Static Features & SVM        & EER = 3.0\%     & -                                                                      & 75          \\ \hline
Sae-Bae \textit{et al.} (2014) \\ \cite{Sae_multiTouch}              & Touchscreen Gestures & Distance between Points & DTW        & EER = 1.58\%     & -                                                                      & 34           \\ \hline
Shen \textit{et al.} (2016) \\ \cite{ChaoSen_TIFS}              & Touchscreen Gestures  &Static Featurs & \begin{tabular}[c]{@{}c@{}}SVM, Random Forest, \\ $k$NN, Neural Networks\end{tabular} & EER $\sim$ 3.0\%     & -                                                                      & 71          \\ \hline
Fierrez \textit{et al.} (2018) \\ \cite{2018_TIFS_Swipe_Fierrez}              & Touchscreen Gestures              &Static Features & SVM, GMM	 & EER = 10.7\%     & -                                                                      & 190          \\ \hline \hline
Sae-Bae \textit{et al.} (2014) \\ \cite{Sae_signature}                 & Handwritten Signatures   &Histogram Static Features & Manhattan Distance                   & EER = 5.04\%           & -                                                                      & 180          \\ \hline
Tolosana \textit{et al.} (2017) \\ \cite{eBioSign_journal}           & Handwritten Signatures  & Dynamic Features & DTW                   & EER = 0.5\%            & EER = 17.9\%                                                                 & 65           \\ \hline \hline
Khan \textit{et al.} (2011) \\ \cite{Khan_graphicalPassword}       & Graphical Passwords   & Predefined Symbols & Exact Match                 & -            & -                                                                 & 100          \\ \hline
Martinez-Diaz \textit{et al.} (2016) \\ \cite{2016_IEEE_THMS_DoodlePass_Marcos}       & Graphical Passwords   & Dynamic Features & DTW, GMM                  & EER = 3.4\%            & EER = 22.1\%                                                                 & 100          \\ \hline \hline
Kutzner \textit{et al.} (2015) \\ \cite{Travieso_2015}            & Handwritten Password  & Static and Dynamic Features & \begin{tabular}[c]{@{}c@{}}Bayes-Nets, $K$Star, \\ $k$NN\end{tabular}        & -                & \begin{tabular}[c]{@{}c@{}}FAR = 10.42\% \\ FRR = unknown\end{tabular} & 32          \\ \hline \hline
Nguyen \textit{et al.} (2017) \\ \cite{Sae_Bae_2017}            & Handwritten Digits  & Dynamic Features & DTW  & -                & EER = 4.84\% & 20          \\ \hline
Tolosana \textit{et al.} (2018) \\ \cite{2018_CVPRW_OTP_Tolosana}    & Handwritten Digits  & Dynamic Features & DTW     & \textbf{-}                & EER = 5.5\%  & 93         \\ \hline
\textbf{Proposed Approach}     & \textbf{Handwritten Digits}  & \textbf{Dynamic Features} & \textbf{DTW, RNNs}     & \textbf{-}                & \textbf{EER = 3.8\%}  & \textbf{93}          \\ \hline
\end{tabular}
\end{adjustbox}
\end{table*}

\section{Related Works}\label{related_works}

\subsection{Handwriting Biometrics and Beyond}
Touch biometrics are becoming a very attractive way to verify users on mobile devices~\cite{2018_TIFS_Swipe_Fierrez, eBioSign_journal}. Table~\ref{comparative_mobileScenarios} summarizes relevant approaches in this area. For each study, we include information related to the verification method, features, classifiers and datasets considered. We also report in Table~\ref{comparative_mobileScenarios} the verification performance for the two impostor scenarios commonly considered in this area: \textit{i)} \textit{imitation attack}, the case in which impostors have some level of information about the user being attacked~\cite{2018_HanbookBioAntiSpoofing_signature_Tolosana}; and \textit{2)} \textit{random attack}, the case in which no information about the user being attacked is known. Note that most algorithms and experimental conditions vary between the listed works, e.g., the amount and type of training and testing data. Therefore, Table~\ref{comparative_mobileScenarios} should be mainly interpreted in general terms to compare different scenarios of use based on touch biometrics, but not individual algorithms. 

In~\cite{Angulo_2011}, Angulo \textit{et al.} evaluated the use of lock pattern dynamic systems for user authentication. Users were asked to draw three different lock patterns a certain number of times (50 trials for each pattern), with each pattern consisting of six dots. Authors considered a total of 11 timing-related features extracted from the finger-in-dot time (i.e., the time in milliseconds from the moment the participant finger touches a dot to the moment the finger is dragged outside the dot area), and the finger-in-between-dots time (i.e., representing the speed at which the finger moves from one dot to the next) achieving results above 10.0\% EER for imitation attacks. In~\cite{Lacharme_2016}, Lacharme \textit{et al.} incorporated biometric dynamic features related to the position of the finger, pressure, finger size and accelerometer sensor to the traditional Android unlock patterns, achieving a final 15.0\% EER for imitation attacks using a matching algorithm based on Hamming Distance. Zezschwitz \textit{et al.} presented in~\cite{deLuca_lockPatterns} a similarity metric for Android unlock patterns to quantify the effective password space of user-defined gestures. The proposed metric was evaluated using 506 user-defined patterns revealing very similar shapes that only differ by simple geometric transformations such as rotation. Consequently, they presented an approach to increase the pattern diversity in order to strengthen user lock patterns.

Other studies have focused on the potential of keystroke biometrics for user authentication on mobile scenarios. In~\cite{deLuca_typing_2}, Buschek \textit{et al.} introduced qualitative aspects like personal expressiveness in order to enhance traditional keystroke biometric systems based on quantitative factors such as error rates and speed. They introduced a dynamic font personalisation framework, TapScript, which adapted a finger-drawn font according to user behavior and context, such as finger placement, device orientation, and position of the user while typing (i.e., walking or sitting) - resulting in a handwritten-looking font. Following their new approach, users were able to distinguish pairs of typists with 84.5\% accuracy and walking/sitting scenarios with 94.8\%. The same authors compared in~\cite{deLuca_typing_1} touch-specific features between three different hand postures (i.e., one-thumb, two-thumb and index finger typing) and evaluation schemes: Gaussian Model without covariance (GM), $k$-Nearest-Neighbours ($k$NN) and Least Squares Anomaly Detection (LSAD). Authors concluded that spatial touch features reduces the Equal Error Rates (EER) by 26.4 - 36.8\% compared to the traditional temporal features.

Biometric verification systems based on touchscreen gestures (i.e., scrolling, zooming and clicking) while using mobile devices in scenarios such as document reading, web surfing or free tasks are gaining a lot of impact nowadays~\cite{Li_2013, Sae_multiTouch, ChaoSen_TIFS, 2018_TIFS_Swipe_Fierrez}. These approaches enable active or continuous authentication schemes, in which the user is transparently authenticated~\cite{Serwadda_BTAS, 2016_SPM_Patel}. Different features and algorithms have been proposed in this field achieving very good results against random attacks. In~\cite{Sae_multiTouch}, the authors proposed a set of 22 multitouch gestures using characteristics of hand and finger movements with an algorithm robust to orientation and translation achieving a final result of 1.58\% EER. In~\cite{2018_TIFS_Swipe_Fierrez}, a set of 100 static features extracted from swipe gestures and systems based on Support Vector Machines (SVM) and Gaussian Mixture Models (GMM) were considered obtaining performances up to 10.7\% EER. Very good results have been also achieved in~\cite{Li_2013, ChaoSen_TIFS} using verification algorithms such as SVM, $k$NN, Random Forest and Neural Networks. 

Handwritten signature is one of the most socially accepted biometrics as it has been used in financial and legal agreements for many years~\cite{FierrezHandbook2008, Plamondon_PAMI_2000, Pirlo_2014, Moises_ACM_2018}, and it also finds applications in mobile scenarios. In~\cite{2018_IEEEAccess_RNN_Tolosana}, the authors explored the use of new algorithms based on RNNs on traditional desktop scenarios for pen-based signature recognition achieving results below 5.0\% EER for imitation attacks. However, a considerable degradation of the system performance with results around 20.0\% EER is obtained for imitation attacks when testing on mobile scenarios using finger touch as input~\cite{Sae_signature, eBioSign_journal}. The main reason for such degradation of the system performance when using finger touch compared to pen-based desktop scenarios is the fact that users tend to modify the way they sign between both approaches, e.g., users who perform their signatures using closed letters with a pen input tend to perform much larger writing executions when using the finger. Besides, users whose signatures are composed of a long name and surname (or two surnames) tend to simplify some parts of their signatures due to the small surface of the screen to sign on. Graphical passwords were studied in~\cite{Khan_graphicalPassword, 2016_IEEE_THMS_DoodlePass_Marcos}. In~\cite{2016_IEEE_THMS_DoodlePass_Marcos}, the authors proposed an approach based on graphical passwords (doodles) achieving final results above 20.0\% EER for imitation attacks. The main reason for such degradation of the system performance lays down on the specific task that the user needs to perform to be authenticated, e.g., doodles were difficult to memorize for most of the users as they didn't use them on a daily basis.

Finally, strongly related to the present work, in~\cite{Travieso_2015, Sae_Bae_2017} the authors proposed the use of handwritten passwords to be authenticated. In~\cite{Travieso_2015}, Kutzner \textit{et al.} asked the users to perform an 8-digit password on the screen of a tablet device. For each handwritten password, a total of 25 static and dynamic features were extracted and tested using many different authentication algorithms. However, the authentication scenario considered in that approach restricts the deployment of the technology in real mobile applications as: \textit{i)} the authors considered a large number of training samples (12), and \textit{ii)} it seems to be only applicable to devices with large screens (such as tablets) as it would be very difficult for the users to perform such a long password (8 digits) on a screen of much smaller size. In~\cite{Sae_Bae_2017}, Nguyen \textit{et al.} evaluated the use of handwritten touch biometrics for PIN-based authentication systems. Their proposed authentication approach overcame some of the drawbacks previously cited as they asked users to draw each digit of the PIN one by one. A final 4.84\% EER was achieved using a biometric system composed of 5 dynamic features and a matcher algotithm based on DTW. Finally, a preliminary study of the work presented here was published in~\cite{2018_CVPRW_OTP_Tolosana}. In that work we performed an initial analysis of the touch biometric system only for OTP authentication schemes. In addition, DTW was the only approach considered for the similarity computation.   

The study presented here extends the preliminary analysis carried out in~\cite{2018_CVPRW_OTP_Tolosana}. The main improvements over~\cite{2018_CVPRW_OTP_Tolosana} are:
\begin{itemize}
\item Our preliminary touch biometric system in~\cite{2018_CVPRW_OTP_Tolosana} based on DTW has been extended by incorporating RNN deep learning architectures. To the best of our knowledge, this is the first work to date that studies recurrent Siamese networks to model handwritten password authentication systems. 
\item Our analysis in~\cite{2018_CVPRW_OTP_Tolosana} studied only OTP schemes. Here we also study PIN authentication, as depicted in Fig.~\ref{diagram_main}. 
\item The system architecture includes 2 new blocks with respect to~\cite{2018_CVPRW_OTP_Tolosana}: \textit{i)} enrolment set, and \textit{ii)} password generation; which are discussed in the text and evaluated experimentally.
\item Sec.~\ref{related_works} has been included to survey and compare advantages and limitations of recent research on touch biometrics for mobile authentication.
\item The results achieved in the present study outperform our initial results presented in~\cite{2018_CVPRW_OTP_Tolosana} with a final 3.8\% EER, a relative improvement of 30.9\% EER compared to~\cite{2018_CVPRW_OTP_Tolosana}. This result outperforms other touch biometric approaches and considers fewer enrolment samples.
\item Sec.~\ref{password_real_systems} has been included to discuss specific details for the deployment of our proposed approach on practical PIN and OTP authentication systems, including various strategies for password generation.

\end{itemize}

\begin{figure*}[tb]
  \centering
    \includegraphics[width=0.6\linewidth]{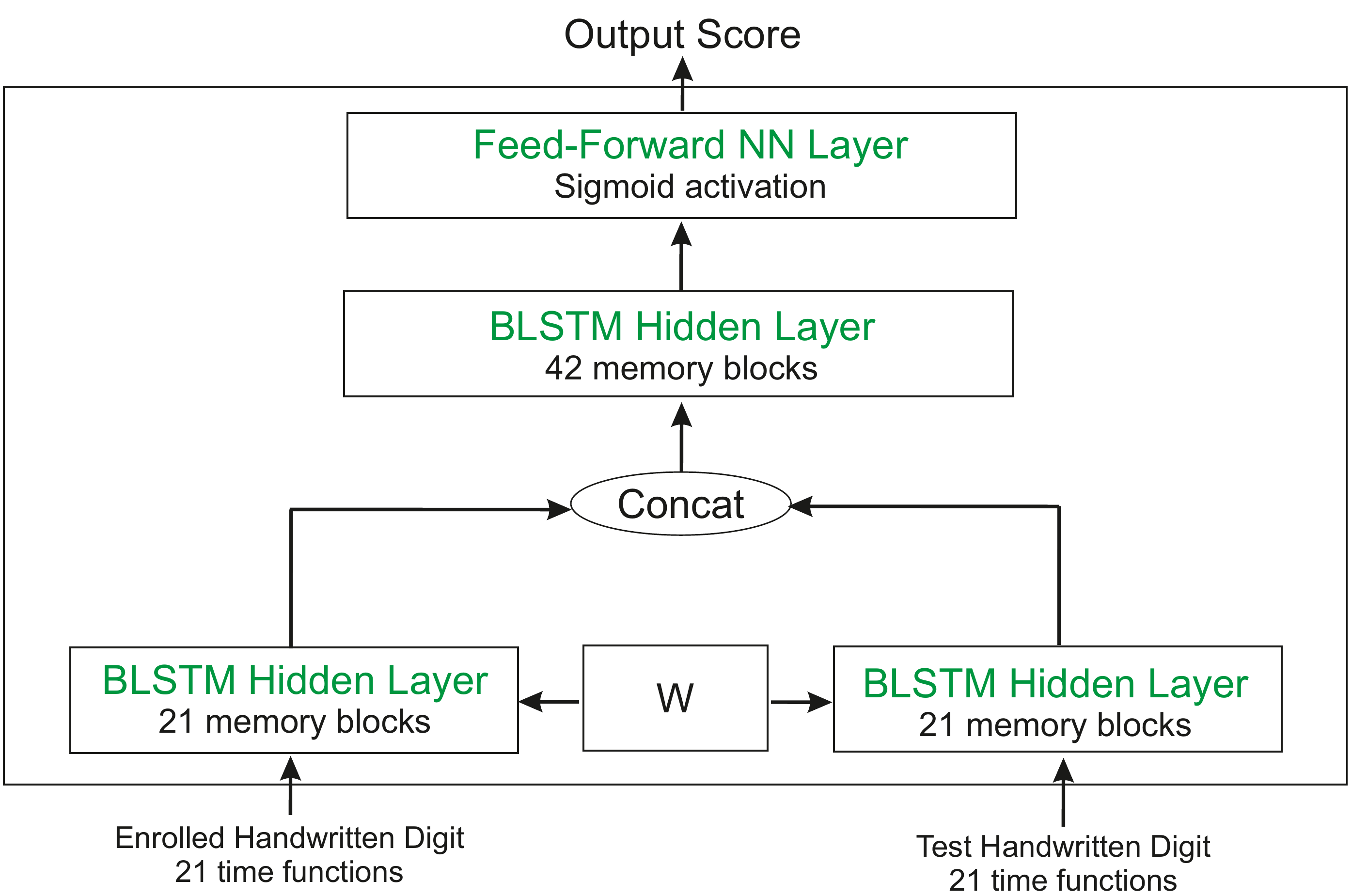}
  \caption{Proposed end-to-end writer-independent BLSTM touch biometric system based on a Siamese architecture.}
  \label{fig:LSTM_configuration_digits}
\end{figure*}

\subsection{Two-Factor Password Authentication}
The incorporation of biometric information on traditional password-based systems can improve the security through a second level of user authentication. Two-factor authentication approaches have been very successful in the last years. These approaches are based on the combination of two authentication stages. For example: \textit{i)} the security system checks that the claimed user introduces its unique password correctly, and \textit{ii)} its behavioral biometric information is used for an enhanced final verification~\cite{Sae_Bae_2017, Luca_2012}. This way the robustness of the security system increases as impostors need more than the traditional password to get access to the system. This approach has been studied in previous works. In~\cite{Angulo_2011}, the authors proposed a two-factor verification system based on timing-related features for dynamic lock patterns, achieving a final average EER of 10.39\% for imitation attacks. A similar two-factor authentication approach was proposed in~\cite{Lacharme_2016} for traditional Android unlock patterns but considering biometric dynamic features related to the position of the finger, pressure, finger size and accelerometer sensor achieving a final 15.0\% EER for imitation attacks. Two-factor authentication approaches have also been expanded to physiological biometric traits. In~\cite{Periocular_2017}, Jenkins \textit{et al.} proposed a system based on features extracted for periocular images acquired using an iPhone 5, achieving very good results for the task of identification.

\begin{table}[tb]
\centering
\caption{Set of time functions considered in this work.}
\begin{adjustbox}{width=0.40\textwidth}
\begin{tabular}{p{1cm}|| p{6cm}}
\hline
\# & Feature \\
\hline \hline
1 & \textit{X}-coordinate: $x_n$  \\
\hline
2 & \textit{Y}-coordinate: $y_n$  \\
\hline
3 & Path-tangent angle: $\theta_n$  \\
\hline
4 & Path velocity magnitude: $v_n$ \\
\hline
5 & Log curvature radius: $\rho_n$ \\
\hline
6 & Total acceleration magnitude: $a_n$ \\
\hline
7-12 & First-order derivative of features 1-6: $\dot{x_n},\dot{y_n},\dot{\theta_n},\dot{v_n},\dot{\rho_n},\dot{a_n}$ \\
\hline
13-14 & Second-order derivative of features 1-2: $\ddot{x_n},\ddot{y_n}$ \\
\hline
15 & Ratio of the minimum over the maximum speed over a 5-samples window: $v^r_n$ \\
\hline
16-17 & Angle of consecutive samples and first-order derivative: $\alpha_n$, $\dot{\alpha_n}$ \\
\hline
18 & Sine: $s_n$ \\
\hline
19 & Cosine: $c_n$ \\
\hline
20 & Stroke length to width ratio over a 5-samples window: $r^5_n$ \\
\hline
21 & Stroke length to width ratio over a 7-samples window: $r^7_n$ \\
\hline
\end{tabular}
\end{adjustbox}
\label{tabla:tablaLocalFeatures}
\end{table}

\section{Touch Biometric System}\label{touch_biometric_system}

\subsection{Digit-based Feature Extraction}\label{SFFS_features_selection}
In this work we evaluate the potential of touch biometric verification systems based on time functions~ \cite{2015_IEEEAccess_InterSign_Tolosana}. Signals captured by the digitizer (i.e., \textit{X} and \textit{Y} spatial coordinates) are used to extract a set of 21 time functions for each numerical digit sample (see Table~\ref{tabla:tablaLocalFeatures}). Information related to pressure, pen angular orientations or pen ups broadly used in other biometric traits such as handwriting and handwritten signature is not considered here as this information is not available in all mobile devices when using the finger touch as input.

Sequential Forward Floating Search (SFFS) algorithm is used for the DTW algorithm in some of the experiments in order to select the best subsets of time functions for each handwritten digit and improve the system performance in terms of EER (\%).

\subsection{Similarity Computation}\label{similarity_computation}

\subsubsection{\textbf{Dynamic Time Warping}}\label{DTW_system}
DTW is used to compare the similarity between genuine and query input samples, finding the optimal elastic match among time sequences that minimizes a given distance measure. Scores are obtained as $score = e^{- D/K}$, where \textit{D} and \textit{K} represent respectively the minimal accumulated distance and the length of the warping path~\cite{Marcos_matching}.

\subsubsection{\textbf{Recurrent Neural Networks}}\label{BLSTM_system}

some of the fields in which RNNs have caused more impact in the last years is in handwriting recognition and writer identification~\cite{online_offline_handwriting, writer_identification_2017}. This study explores the potential of RNNs for the task of handwritten passwords. In particular, we consider an adaptation of our original RNN systems proposed in~\cite{2018_IEEEAccess_RNN_Tolosana} for the task of on-line handwritten signature verification. In that work we proposed RNN systems based on a Siamese architecture. The main goal was to learn a dissimilarity metric from data minimizing a discriminative cost function that drives the dissimilarity metric to be small for pairs of genuine samples from the same subject, and higher for pairs of samples coming from different subjects. Both Long Short-Term Memory (LSTM) and Gated Recurrent Unit (GRU) systems were studied in~\cite{2018_IEEEAccess_RNN_Tolosana}, outperforming a state-of-the-art DTW system in challenging scenarios where skilled forgeries were considered. In addition, in~\cite{2018_IEEEAccess_RNN_Tolosana} we also studied bidirectional schemes (i.e., BLSTM and BGRU), which allow access to future context, achieving much better results compared to the original schemes that only had access to past and present contexts. 

\begin{figure*}[tb]
\centering
\subfigure[]{\label{acquisition}
\includegraphics[width=0.175\linewidth]{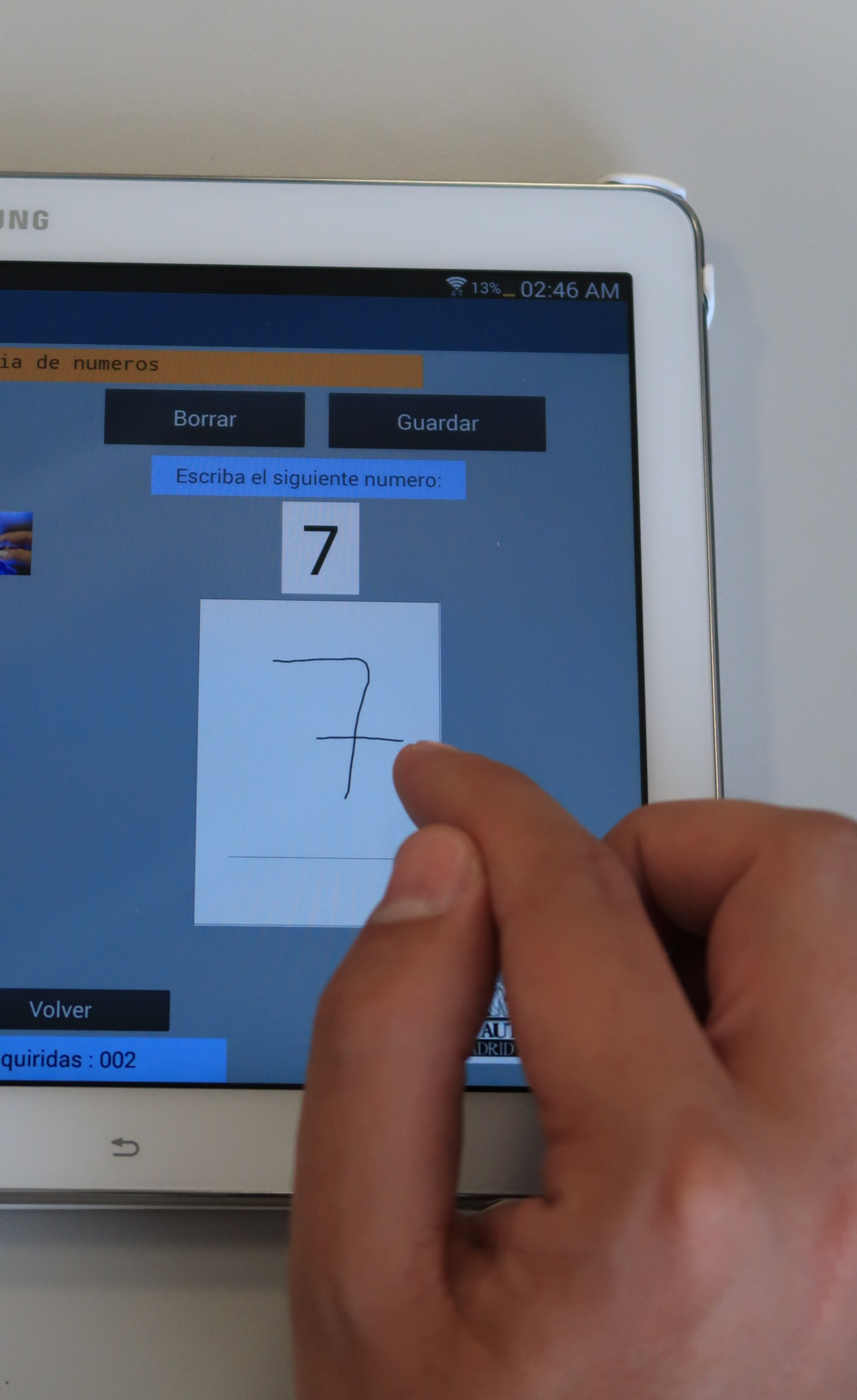}}
\hspace{0.3cm}
\subfigure[]{\label{digit_0}
\includegraphics[width=0.19\linewidth]{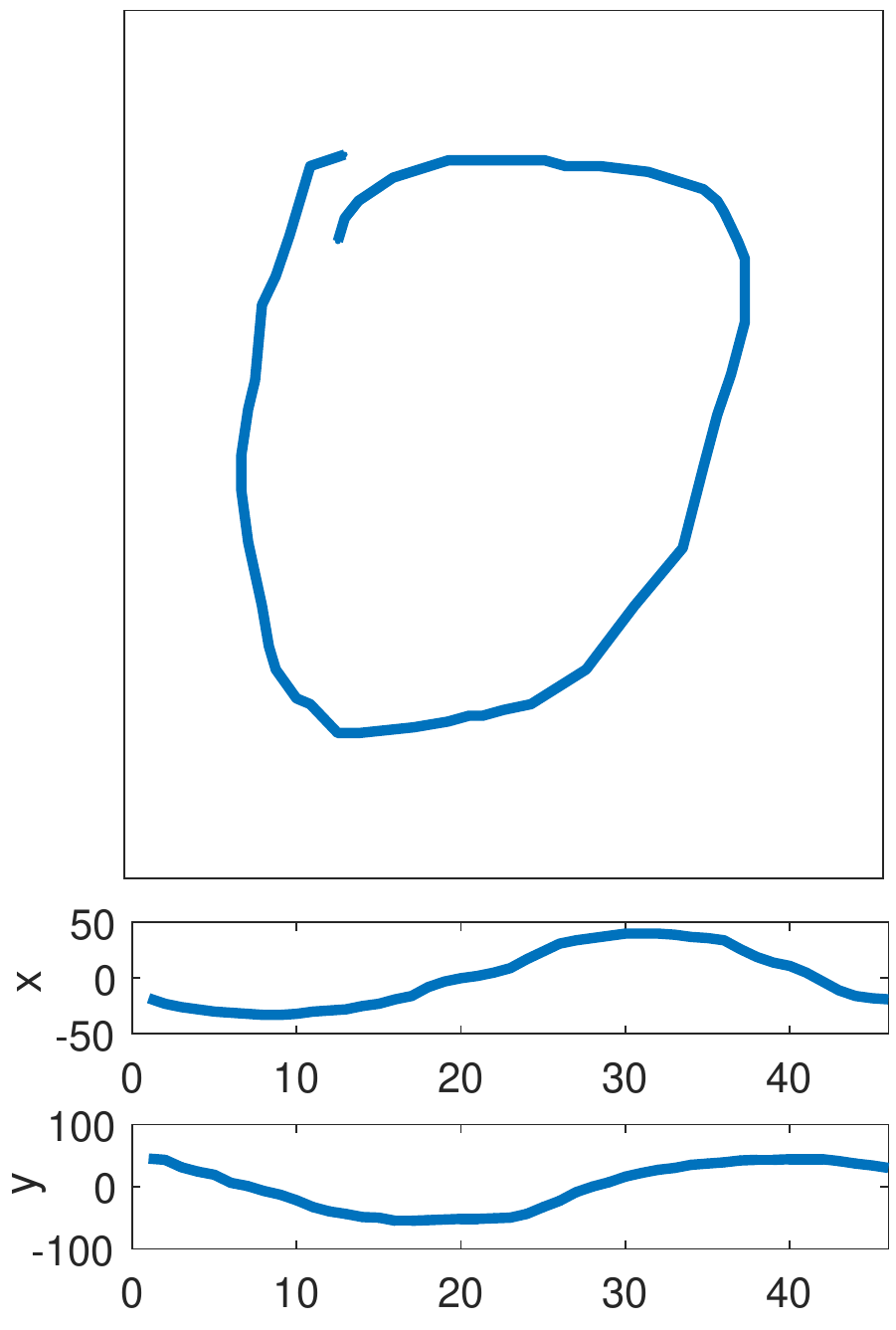}}
\hspace{0.3cm}
\subfigure[]{\label{digit_2}
\includegraphics[width=0.195\linewidth]{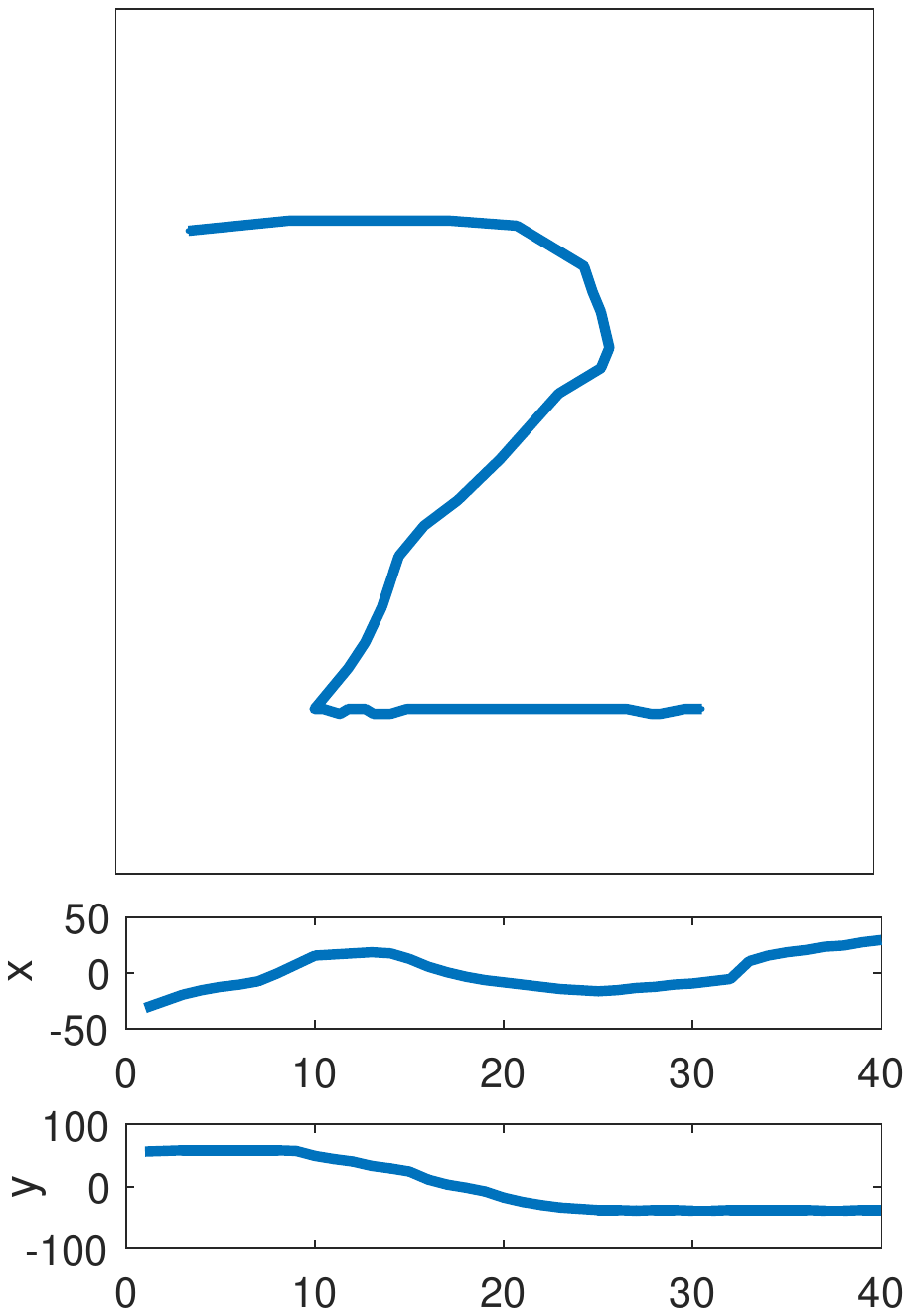}}
\subfigure[]{\label{digit_7}
\hspace{0.3cm}
\includegraphics[width=0.19\linewidth]{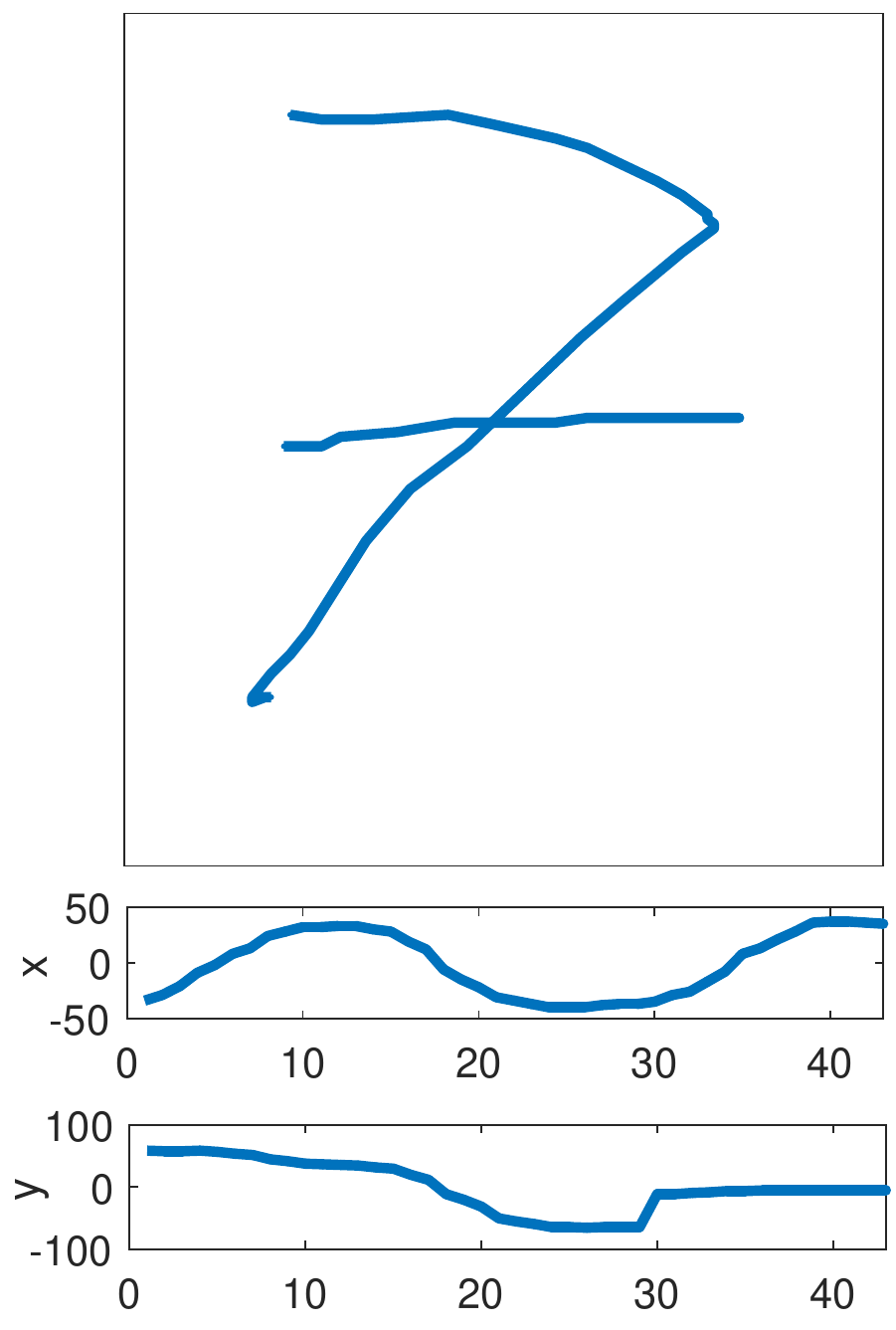}}
\caption{(a) Acquisition setup. (b-d) examples of different handwritten numerical digits of the e-BioDigit database. \textit{X} and \textit{Y} denote horizontal and vertical position versus the time samples.} \label{example_digits}
\end{figure*}

In this study we adapt the original BLSTM system proposed in~\cite{2018_IEEEAccess_RNN_Tolosana} to handwritten passwords for touchscreen biometrics. To the best of our knowledge, this is the first work to date that studies recurrent Siamese networks to model handwritten password authentication systems. Fig.~\ref{fig:LSTM_configuration_digits} shows our proposed end-to-end writer-independent BLSTM touch biometric system based on a Siamese architecture. For the input of the system, we feed the network with as much information as possible, i.e., all 21 time functions per digit. The first layer is composed of two BLSTM hidden layers with 21 memory blocks each, sharing the weights between them. The outputs of the first two parallel BLSTM hidden layers are concatenated and serve as input to the second layer, which corresponds to a BLSTM hidden layer with 42 memory blocks. Finally, a feed-forward neural network layer with a sigmoid activation is considered, providing an output score for each pair of digits. It is important to highlight that our approach is trained to distinguish between genuine and impostor patterns from all numerical digits and users. Thus, we just train one writer-independent system for all digits and users through a development dataset.

\section{Database e-BioDigit}{\label{e_BioDigit_description}}
The new e-BioDigit database was captured in order to perform the experimental work included in this article. This database comprises on-line handwritten numerical digits from 0 to 9 acquired using a Samsung Galaxy Note 10.1 general purpose tablet. This device has a 10.1-inch LCD display with a resolution of 1280$\times$800 pixels.

Regarding the acquisition protocol, subjects had to perform handwritten numerical digits from 0 to 9, one at a time. The acquisition setup and some examples of the handwritten numerical digits of the e-BioDigit database are depicted in Fig.~\ref{example_digits}. Additionally, samples were collected in two sessions with a time gap of at least three weeks between them in order to consider inter-session variability, very important for behavioral biometric traits~ \cite{galbally13PONEagingSignature}. For each session, users had to perform a total of 4 numerical sequences from 0 to 9 using the finger as input. Therefore, there are a total of 8 samples per numerical digit and user. 

The software for capturing handwritten numerical digits was developed in order to minimize the variability of the user during the acquisition process. A rectangular area with a writing surface size similar to a 5-inch screen smartphone was considered, see Fig.~\ref{acquisition}. A horizontal line was represented on top of the drawing rectangular area, including two buttons ``OK'' and ``Cancel'' to press after writing if the sample was good or bad respectively. 

The database comprises a total of 93 users. Regarding the age distribution, the majority of the subjects (85.0\%) are between 17 and 27 years old, as the database was collected in a university environment (36.6\% between 17 and 21). Regarding the gender, 66.7\% of the subjects were males and 33.3\% females whereas for the handedness distribution, 89.2\% of the population was righthanded.

\begin{figure*}[tb]
\centering
\subfigure[User A, sample 1]{\label{digit_7_u1_1}
\includegraphics[width=0.17\linewidth]{digito7_complejo_1_rl}}
\hspace{0.3cm}
\subfigure[User A, sample 2]{\label{digit_7_u1_2}
\includegraphics[width=0.171\linewidth]{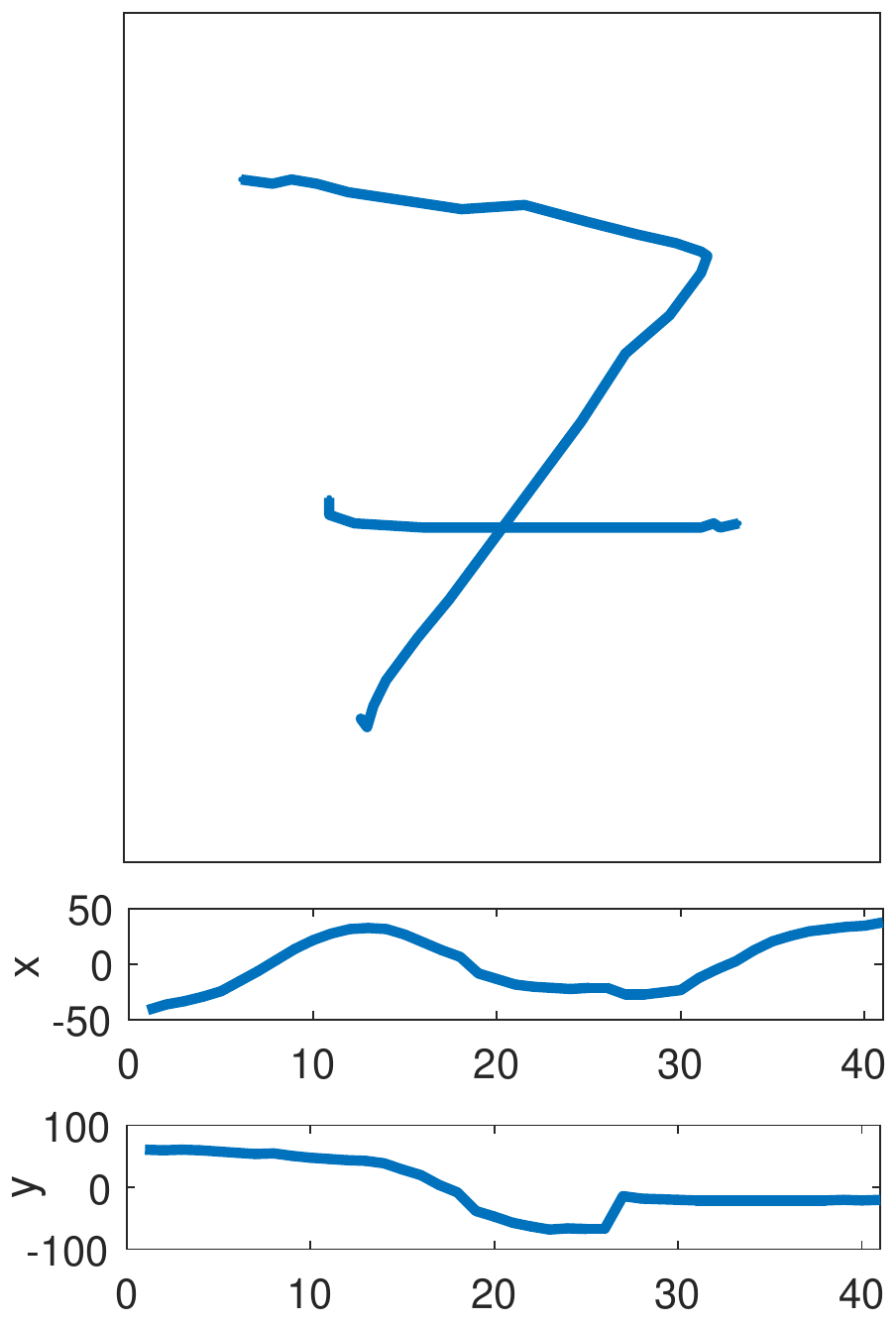}}
\subfigure[User B, sample 1]{\label{digit_7_u1_2}
\hspace{0.3cm}
\includegraphics[width=0.176\linewidth]{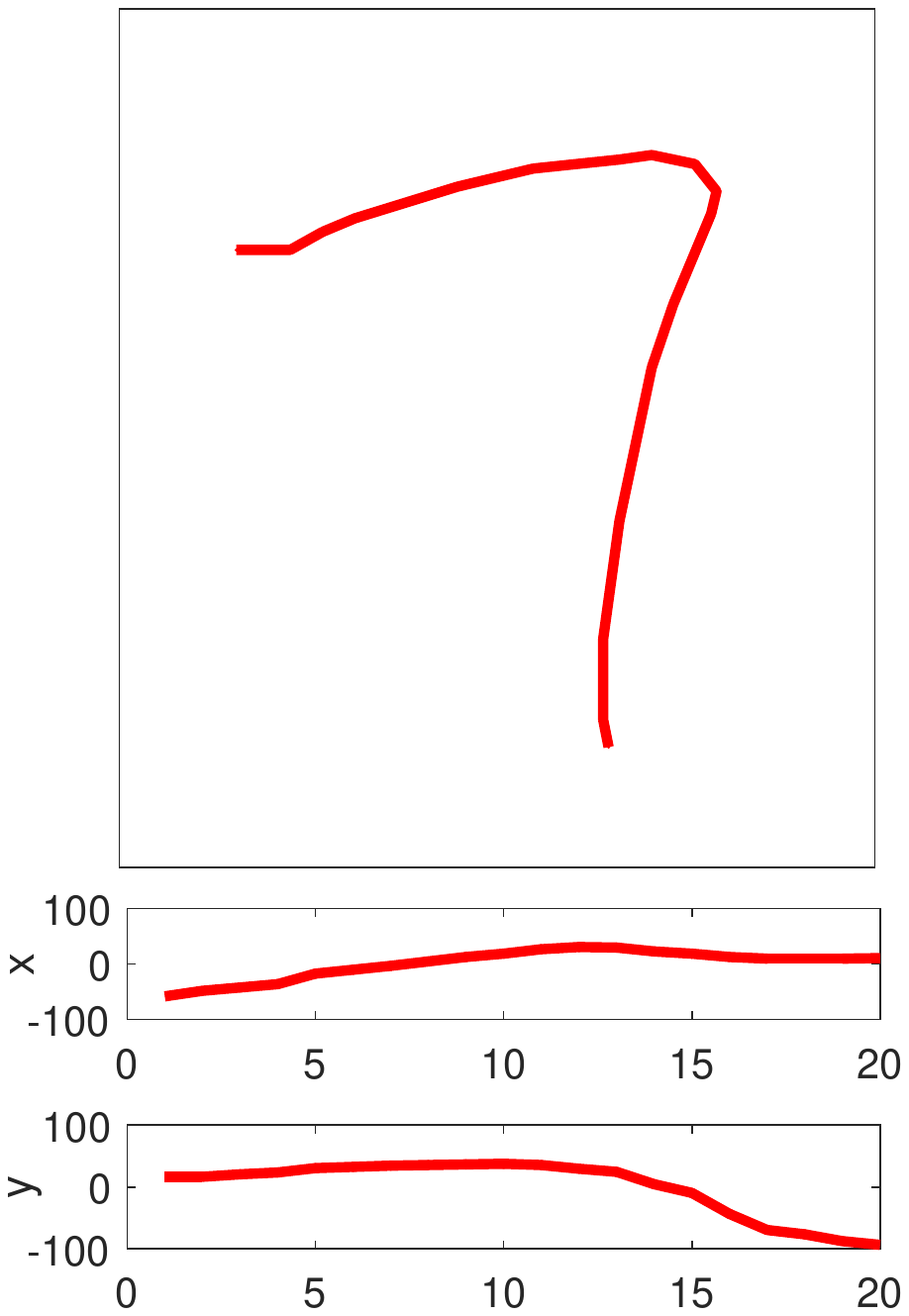}}
\subfigure[User B, sample 2]{\label{digit_7_u1_2}
\hspace{0.3cm}
\includegraphics[width=0.17\linewidth]{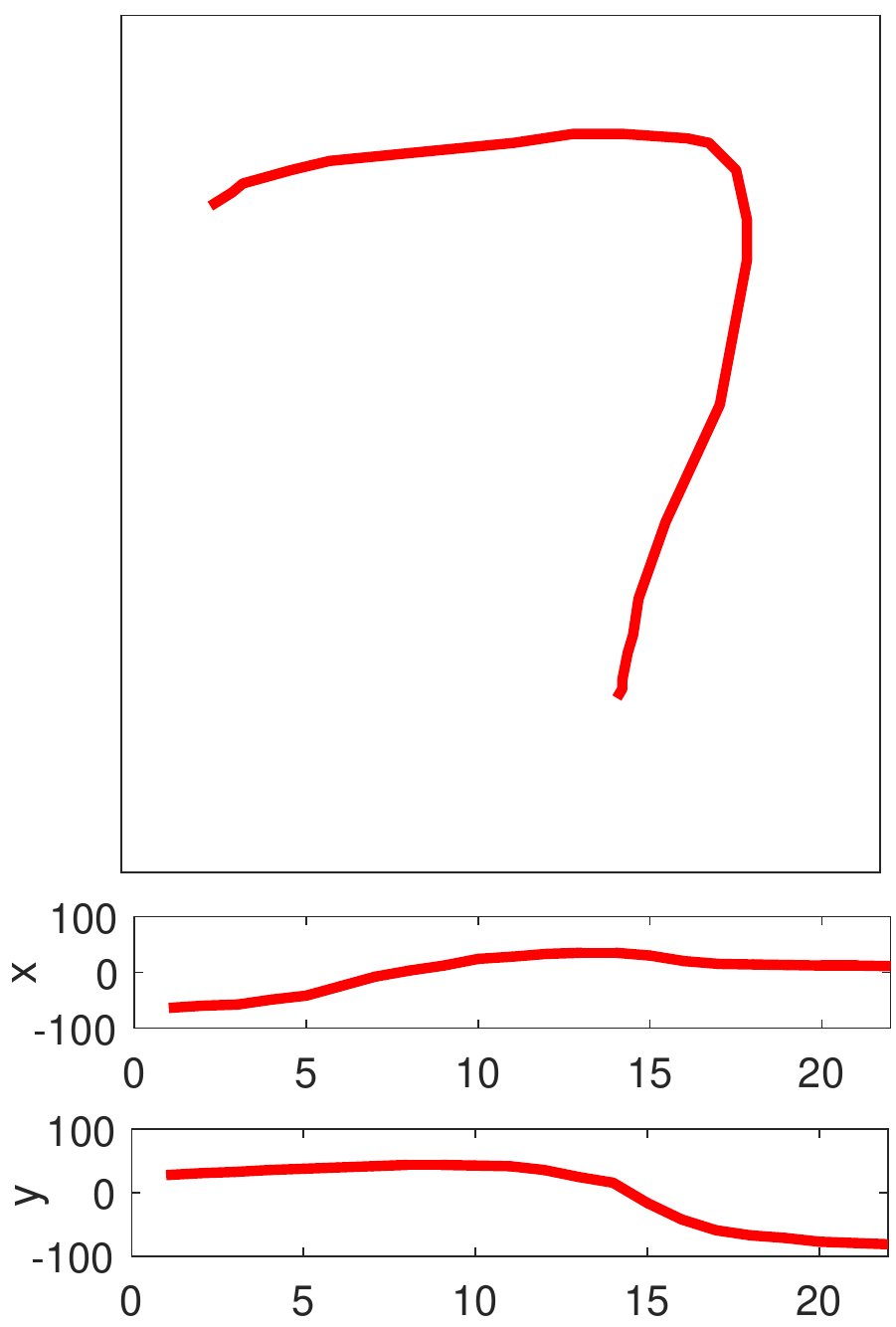}}
\caption{Examples of the digit 7 performed by two different users.} \label{exp1_digit7}
\end{figure*}

\section{Experimental Protocol}\label{experimental_protocol}
The experimental protocol designed in this study intends to cover all details of the two following main modules of our proposed password-based touch biometric system (see Fig.~\ref{diagram_main}):

\begin{itemize}
\item \textbf{Enrolment Set:} When designing biometric authentication systems for real applications, there are usually two conflicting factors: \textit{i)} the amount of data requested to the user during the enrolment, and \textit{ii)} the security level provided by the biometric system. From the point of view of the security system, it seems clear that the ideal case would be to have as much information of the user as possible. However, in most real scenarios, the feasibility and success depend on the development of user-friendly applications. 

This aspect has shown to be crucial for different tasks such as the handwritten signature. In~ \cite{2015_WIFS_SignatureHMMUpdate_RubenT}, we evaluated this effect using statistical systems based on HMM and GMM, achieving an absolute improvement of 11.7\% EER when training the user models with 41 genuine signatures instead of just 4. In this work, we analyze the intra-user variability on this new authentication scenario and perform a complete analysis of how the biometric system performance changes with the number of enrolment samples acquired per digit.

\item \textbf{Password Generation:} The selection of a password that is robust enough for a specific application is a key factor. The number of digits that comprise the password depends on the scenario and level of security considered in the final application. For example, for everyday applications such as Facebook or Gmail, it is not reasonable from the point of view of the users to memorize passwords composed of 12 digits. Additionally, OTP-based systems could request longer passwords compared to PIN-based systems as users do not have to memorize them, i.e., the security system is in charge of selecting and providing the password to the user. 

In this experimental work we evaluate the robustness of handwritten passwords regarding the three following features: \textit{i)} which digits better discriminate users, \textit{ii)} whether repetitions of the same numerical digits in a password can help to discriminate users or not, and \textit{iii)} the length of the password. For short passwords (i.e., fewer than 6 digits), this analysis is carried out performing all possible digit combinations, whereas for longer passwords, the SFFS algorithm is used to select the best digit combinations due to the high cost of performing all possible comparisons.
\end{itemize}

\begin{table}[tb]
\centering
\caption{Time functions selected for the Baseline System.}
\begin{tabular}{c||c}
\hline
\textbf{\#} & Time-function description \\
\hline \hline
1 & \textit{X}-coordinate: $x_n$  \\
\hline
2 & \textit{Y}-coordinate: $y_n$  \\
\hline
7-8 & First-order derivate of features 1-2: $\dot{x_n},\dot{y_n}$ \\
\hline
13-14 & Second-order derivate of features 1-2: $\ddot{x_n},\ddot{y_n}$ \\
\hline
\end{tabular}
\label{BaselineSystem_timeFunctions}
\end{table}

In order to perform a complete analysis of these two modules, the e-BioDigit database is divided into development (the first 50 users) and evaluation (the remaining 43 users) datasets. 

For the development of our proposed handwritten touch biometric systems, $N$ genuine signatures (up to 4) from the first session can be used as enrolment samples, whereas the 4 remaining genuine samples from the second session are used for testing. Impostor scores are obtained by comparing the $N$ enrolment samples with one genuine sample of each of the remaining users (simulating this way the imitation attack in which the impostor knows the password).

\begin{table*}[tb]
\centering
\caption{System performance as EER(\%) of each numerical digit for the \textbf{1vs1} case on the evaluation dataset.}
\label{experiment1_1vs1}
\begin{adjustbox}{width=0.7\textwidth}
\begin{tabular}{c||cccccccccc}
\cline{2-11}
                           & \multicolumn{10}{c}{Numerical Digit}                               \\ \cline{2-11} 
                           & 0    & 1    & 2    & 3    & 4    & 5    & 6    & 7    & 8    & 9    \\ \hline \hline
\multicolumn{1}{c||}{DTW Baseline System} & 34.9 & 32.3 & 32.8 & 35.0 & 23.5 & 24.4 & 36.9 & \textbf{22.5} & 26.0 & 29.6 \\ \hline 
\multicolumn{1}{c||}{DTW Adapted System} & 33.0 & 34.0 & 30.9 & 32.3 & 22.0 & \textbf{21.7} & 33.6 & 21.8 & 21.8 & 27.0 \\ \hline
\multicolumn{1}{c||}{BLSTM System} & 32.8 & 30.8 & 32.8 & 32.3 & 26.2 & \textbf{19.6} & 35.2 & 28.5 & 21.7 & 23.8 \\ \hline

\end{tabular}
\end{adjustbox}
\end{table*}

\begin{table*}[!]
\centering
\caption{System performance as EER(\%) of each numerical digit for the \textbf{4vs1} case on the evaluation dataset.}
\label{experiment1_4vs1}
\begin{adjustbox}{width=0.7\textwidth}
\begin{tabular}{c||cccccccccc}
\cline{2-11}
                           & \multicolumn{10}{c}{Numerical Digit}                               \\ \cline{2-11} 
                           & 0    & 1    & 2    & 3    & 4    & 5    & 6    & 7    & 8    & 9    \\ \hline \hline
\multicolumn{1}{c||}{DTW Baseline System} & 33.1 & 28.5 & 30.2 & 32.6 & \textbf{18.0} & 20.3 & 36.6 & 19.2 & 22.7 & 25.0 \\ \hline 
\multicolumn{1}{c||}{DTW Adapted System} & 31.4 & 33.1 & 27.9 & 29.7 & 19.2 & \textbf{16.9} & 29.7 & 20.3 & 18.6 & 23.3 \\ \hline
\multicolumn{1}{c||}{BLSTM System} & 31.4 & 27.9 & 31.4 & 26.2 & 24.4 & \textbf{17.4} & 35.4 & 24.4 & 18.0 & 20.9 \\ \hline
\end{tabular}
\end{adjustbox}
\end{table*}

For the evaluation of our proposed touch biometric system, different scenarios are generally considered regarding the number of available enrolment samples per user (i.e., $N$vs1), in which the final score is performed as the average score of $N$ one-to-one comparisons. In addition, in case of using passwords composed of several digits, the final score is produced after averaging the different one by one digit score comparisons. 

It is important to highlight that the inter-session variability problem is also considered in the experimental protocol carried out in this work as genuine digit samples from different sessions are used as enrolment and testing samples respectively. This effect has proven to be very important for many behavioral biometric traits such as the case of the handwritten signature~\cite{galbally13PONEagingSignature}.

\section{Experimental Results}\label{experimental_results}

\subsection{\textbf{One-Digit Analysis.}}
This section analyzes the potential of each numerical digit (i.e., from 0 to 9) for the task of user authentication. We consider three different systems: \textit{i)} a baseline DTW system, \textit{ii)} an adapted DTW considering feature selection, and \textit{iii)} a system based on RNNs. 

Experimental results on the evaluation dataset for these three systems are shown in Table~\ref{experiment1_1vs1} and~\ref{experiment1_4vs1} in terms of EER (\%) for the cases of 1vs1 and 4vs1 comparisons, respectively.

\subsubsection{\textbf{DTW Baseline System}}
In order to provide an easily reproducible framework, we first consider a baseline system based on DTW with the same fixed time functions for all numerical digits. Table~\ref{BaselineSystem_timeFunctions} shows the time functions selected, which are commonly used as baseline in other biometric traits such as the handwritten signature~\cite{eBioSign_journal, iet:/content/journals/10.1049/iet-bmt.2013.0044}.  

\begin{figure*}[htb]
  \centering
    \includegraphics[width=0.9\linewidth]{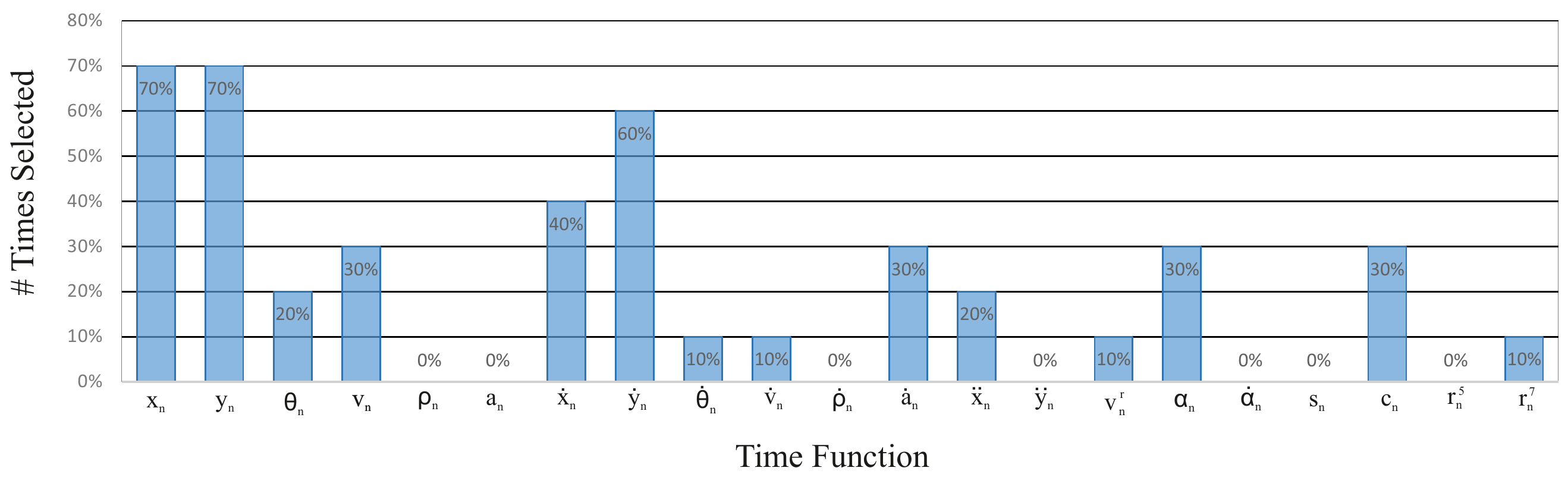}
  \caption{Histogram of functions selected by SFFS for our DTW Adapted System. Functions described in Table~\ref{tabla:tablaLocalFeatures}.}
  \label{exp2_timefunctions}
\end{figure*}

Analyzing the first rows of Tables~\ref{experiment1_1vs1} and~\ref{experiment1_4vs1} we can see how very good authentication results are obtained by the DTW Baseline System taking into account that we only consider one digit and the same time functions for all numerical digits. 

Analyzing in Table~\ref{experiment1_1vs1} the extreme scenario of having just one available digit sample during the enrolment (1vs1), the numerical digit 7 achieves the best result with 22.5\% EER. In addition, other numerical digits such as 4 or 5 achieve similar results with EERs below 25.0\%. This first experiment puts in evidence the discriminative power of each handwritten digit. Fig.~\ref{exp1_digit7} shows examples of the digit 7 performed by two different users in order to observe the low intra- and high inter-user variability of this number. This effect is produced as different users tend to perform a specific digit in a different way, i.e., starting from a different stroke of the digit or even removing some of them such as the crossed horizontal stroke of the number 7. 

Analyzing in Table~\ref{experiment1_4vs1} the scenario of using four enrolment samples (4vs1), an average absolute improvement of 3.2\% EER is achieved compared to the 1vs1 scenario showing the importance of acquiring more than one sample during the enrolment stage, if possible. For this scenario, the digit 4 achieves the best result with 18.0\% EER.

\subsubsection{\textbf{DTW Adapted System}}
We now apply SFFS over the development dataset in order to enhance the DTW touch biometric system through the selection of specific time functions for each handwritten digit. Fig.~\ref{exp2_timefunctions} shows the number of times each time function is selected in our DTW Adapted System from the 21 total time functions described in Table~\ref{tabla:tablaLocalFeatures}. In general, we can highlight the importance of $x_n$, $y_n$ time functions as they are selected for 70\% of the numerical digits. In addition, time functions $\dot{x_n}, \dot{y_n}$ related to \textit{X} and \textit{Y} time derivatives seem to be very important as they are selected for near half of the digits. Other time functions such as $\rho_n$, $\dot{\rho_n}$, $\dot{\alpha_n}$ and $s_n$ related to geometrical aspects of the numerical digits are proven not to be very useful to discriminate between genuine and impostor users.

The second rows of Tables~\ref{experiment1_1vs1} and~\ref{experiment1_4vs1} show the results achieved for each digit using our DTW Adapted System over the evaluation dataset for both 1vs1 and 4vs1 cases, respectively. In general, better results are achieved compared to the DTW Baseline System. Analyzing the 1vs1 scenario, our DTW Adapted System achieves an average absolute improvement of 2.0\% EER, being the numerical digit 5 the one that provides the best result with a 21.7\% EER. Analyzing the 4vs1 scenario, our DTW Adapted System achieves an average absolute improvement of 1.6\% EER, being again the numerical digit 5 the one that achieves the best result with a 16.9\% EER. These results put in evidence the importance of considering different time functions for each digit in order to develop more robust biometric authentication systems against attacks. 

\begin{table*}[tb]
\centering
\caption{Evolution of the system performance in terms of EER (\%) on the evaluation dataset. The best system performance achieved and the corresponding handwritten digits selected are shown on top and bottom of each cell respectively.}
\label{exp3_enrolment_length}
\begin{adjustbox}{width=0.95\textwidth}
\begin{tabular}{cc||cccccccc}
\cline{3-10}
                                                                                                      &   & \multicolumn{8}{c}{\# Digits that comprise the password}                                                                                                                                                                                                                                                                       \\ \cline{3-10} 
                                                                                                      &   & 1                                                & 2                                                  & 3                                                    & 4                                                      & 5 & 6 & 7 & 8                                                                                                  \\ \hline \hline
\multicolumn{1}{c|}{\multirow{4}{*}{}} & 1 & \begin{tabular}[c]{@{}c@{}}21.7\\ $[5]$ \end{tabular} & \begin{tabular}[c]{@{}c@{}}14.0\\ $[5, 8]$\end{tabular} & \begin{tabular}[c]{@{}c@{}}11.6\\ $[5, 7, 9]$\end{tabular} & \begin{tabular}[c]{@{}c@{}}11.6\\ $[1, 5, 7, 9]$\end{tabular} & \begin{tabular}[c]{@{}c@{}}9.3\\ $[2, 5, 6, 7, 8]$\end{tabular} & \begin{tabular}[c]{@{}c@{}}8.5\\ $[2, 3, 5, 6, 7, 8]$\end{tabular}  & \begin{tabular}[c]{@{}c@{}}8.5\\ $[1, 2, 3, 5, 6, 7, 8]$\end{tabular} & \begin{tabular}[c]{@{}c@{}}8.5\\ $[2, 3, 4, 5, 6, 7, 8, 9]$\end{tabular} \\ \cline{2-10} 
\multicolumn{1}{c|}{\multirow{4}{*}{\begin{tabular}[c]{@{}c@{}}\# Enrolment\\ samples\end{tabular}}}                                                                                & 2 & \begin{tabular}[c]{@{}c@{}}18.6\\ $[5]$\end{tabular} & \begin{tabular}[c]{@{}c@{}}11.6\\ $[5, 8]$\end{tabular} & \begin{tabular}[c]{@{}c@{}}9.3\\ $[2, 5, 8]$\end{tabular}  & \begin{tabular}[c]{@{}c@{}}7.4\\ $[2, 5, 8, 9]$\end{tabular}  & \begin{tabular}[c]{@{}c@{}}7.3\\ $[1, 2, 5, 7, 9]$\end{tabular} & \begin{tabular}[c]{@{}c@{}}4.6\\ $[2, 5, 6, 7, 8, 9]$\end{tabular} & \begin{tabular}[c]{@{}c@{}}4.6\\ $[1, 2, 3, 5, 7, 8, 9]$\end{tabular} & \begin{tabular}[c]{@{}c@{}}4.6\\ $[1, 2, 3, 4, 5, 6, 7, 8]$\end{tabular} \\ \cline{2-10} 
\multicolumn{1}{c|}{}                                                                                & 3 & \begin{tabular}[c]{@{}c@{}}16.3\\ $[5]$\end{tabular} & \begin{tabular}[c]{@{}c@{}}9.5\\ $[2, 8]$\end{tabular}  & \begin{tabular}[c]{@{}c@{}}7.4\\ $[1, 2, 8]$\end{tabular}  & \begin{tabular}[c]{@{}c@{}}5.9\\ $[2, 5, 8, 9]$\end{tabular}  & \begin{tabular}[c]{@{}c@{}}4.7\\ $[1, 2, 5, 8, 9]$\end{tabular} & \begin{tabular}[c]{@{}c@{}}4.6\\ $[1, 2, 3, 5, 8, 9]$\end{tabular} & \begin{tabular}[c]{@{}c@{}}\textbf{3.8}\\ $\textbf{[1, 2, 3, 4, 5, 8, 9]}$\end{tabular} & \begin{tabular}[c]{@{}c@{}}4.6\\ $[0, 1, 2, 3, 4, 5, 7, 8]$\end{tabular} \\ \cline{2-10} 
\multicolumn{1}{c|}{}                                                                                & 4 & \begin{tabular}[c]{@{}c@{}}16.9\\ $[5]$\end{tabular} & \begin{tabular}[c]{@{}c@{}}11.6\\ $[5, 8]$\end{tabular} & \begin{tabular}[c]{@{}c@{}}7.0\\ $[7, 8, 9]$\end{tabular}  & \begin{tabular}[c]{@{}c@{}}6.1\\ $[5, 7, 8, 9]$\end{tabular}  & \begin{tabular}[c]{@{}c@{}}4.7\\ $[1, 5, 7, 8, 9]$\end{tabular} & \begin{tabular}[c]{@{}c@{}}4.6\\ $[1, 2, 5, 7, 8, 9]$\end{tabular} & \begin{tabular}[c]{@{}c@{}}4.3\\ $[1, 2, 3, 5, 7, 8, 9]$\end{tabular} & \begin{tabular}[c]{@{}c@{}}4.8\\ $[0, 1, 2, 3, 4, 5, 7, 8]$\end{tabular} \\ \hline
\end{tabular}
\end{adjustbox}
\end{table*}
   
\subsubsection{\textbf{BLSTM System}}
We now explore the potential of state-of-the-art deep learning technology applied to our touch biometric data. Our proposed end-to-end writer-independent BLSTM system is trained using only the 50 users of the development dataset. Samples from all numerical digits (i.e., from 0 to 9) and development users are considered together during training as we intend to distinguish between genuine and impostor handwritten digit samples regardless of the user and the numerical digit. This approach resulted in better generalisation results compared to the case of training one system per numerical digit. Therefore, our BLSTM system is trained considering two different cases: \textit{i)} pairs of genuine digit samples drawn by the same user, and \textit{ii)} pairs of genuine and impostor digit samples, one performed by the claimed user and the other one by an impostor. For each case there are a total of 4 train samples $\times$ 4 test samples $\times$ 10 numerical digits $\times$ 50 users $\simeq$ 8,000 comparisons, having the same number of genuine and impostor comparisons. Our BLSTM System has been implemented under Keras using Tensorflow as back-end, with a NVIDIA GeForce RTX 2080 Ti GPU. Adam optimizer is considered with a learning rate of 0.001 and a loss function based on binary cross-entropy. 

The third rows of Tables~\ref{experiment1_1vs1} and~\ref{experiment1_4vs1} show the results achieved for each digit using our BLSTM System over the evaluation dataset for both 1vs1 and 4vs1 cases, respectively. In general, better results are achieved compared to the DTW Baseline System. Analyzing the 1vs1 scenario, our BLSTM System achieves an average absolute improvement of 1.4\% EER, being the numerical digit 5 the one that provides the best result with a 19.6\% EER. Analyzing the 4vs1 scenario, our BLSTM System achieves an average absolute improvement of 0.9\% EER, being again the numerical digit 5 the one that achieves the best result with a 17.4\% EER.

Finally, we compare our BLSTM System to the DTW Adapted System. In general, very similar results have been achieved for both authentication systems. For example, analyzing the 1vs1 case in Table~\ref{experiment1_1vs1}, the BLSTM System has outperformed the DTW Adapted System for the 50\% of the numerical digits (i.e., 0, 1, 5, 8, and 9), proving the potential of deep learning technologies even when just a single enrolment sample is considered. Despite these improvements, the DTW Adapted System outperforms slightly the BLSTM System in general, achieving an average absolute improvement of 0.5\% and 0.7\% EER for the 1vs1 and 4vs1 cases, respectively.

\subsection{\textbf{Digit Combinations}}\label{longitud_enrolment_analysis}
This section explores the robustness of our proposed approach when increasing the length of the password and also the number of available enrolment samples. The DTW Adapted System is considered in this analysis as it has outperformed the other systems studied. Regarding the type of digits that comprises the password, repetitions of the same numerical digits are allowed. However, the number of repetitions is restricted to 4, e.g., ``2 5 8 8 8 8". The reason for this limitation is motivated due to only 4 samples were acquired per digit during the second session of the e-BioDigit database. Table~\ref{exp3_enrolment_length} shows the evolution of the system performance in terms of EER (\%) on the evaluation dataset when increasing the length of the handwritten password (from 1 to 8 digits) and also the number of available enrolment samples (from 1 to 4). 

First, we analyze how the length of the handwritten password affects the system performance. In general, a considerable system performance improvement is achieved when adding more handwritten digits to the password. For example, for the case of having just one enrolment sample per user (1vs1), a password that is composed of just two handwritten digits achieves a 14.0\% EER, an absolute improvement of 7.7\% EER compared to the case of using a password with just one digit. This result is further improved when increasing the number of handwritten digits of the password with a final 8.5\% EER for the case of considering a 6-digit password. However, there seems to exist a limit in the system performance improvement with the number of digits that comprise the password. In our experiments, the best results are obtained for passwords with a length of 6 and 7 digits. 

Now, we analyze the effect of the number of available enrolment samples on the system performance. In general, the system performance improves with the number of enrolment samples. For example, for the case of having just one enrolment sample and a password composed of just one digit, the biometric system achieves a 21.7\% EER. This result is further improved when increasing the number of enrolment samples to 4, achieving a final value of 16.9\% EER, an absolute improvement of 4.8\% EER. However, there seems to exist a limit in the system performance improvement with the number of enrolment samples. In our experiment, very similar results are obtained when considering 3 or 4 enrolment samples, achieving a final value of 3.8\% EER when considering 3 enrolment samples and a handwritten password of 7 digits. This interesting finding is different compared to other behavioral biometric traits such as the handwritten signature as the system performance keeps improving even with large number of enrolment samples \cite{2015_WIFS_SignatureHMMUpdate_RubenT}. This effect may be due to the lower intra-user variability of our proposed touch biometric approach compared to other behavioral biometrics as well as the DTW similarity computation algorithm considered.

Finally, we pay attention to the content and the number of possible combinations of the best handwritten passwords using our proposed touch biometric system so as to achieve the best system performance. Table~\ref{exp3_enrolment_length} indicates in the bottom of each cell the best handwritten digits selected but not their order, as the final score of our proposed touch biometric system is produced after averaging the different one by one digit score comparisons. Therefore, for the case of having a password comprised of $n$ digits, there are a total of $n!$ possible password combinations (note that in our experiments we did not have any case of repetitions of digits achieving the best results).

\subsection{\textbf{Comparison to the State of the Art}}\label{state_of_the_art_comparison}
Our proposed approach is now compared to other state-of-the-art biometric authentication approaches described in Table~\ref{comparative_mobileScenarios}. In order to perform a fair analysis, we compare our proposed approach to all studies that consider the same type of impostors, i.e., imitation attacks. 

In general, our proposed approach achieves better results than other touch biometric approaches. For the case of lock pattern dynamic systems~\cite{Angulo_2011, Lacharme_2016}, the best system performance reported was an average 10.39\% EER. Our proposed approach also outperforms other biometric methods such as the handwritten signature or graphical passwords~\cite{eBioSign_journal, 2016_IEEE_THMS_DoodlePass_Marcos}. In~\cite{eBioSign_journal}, the authors proposed handwritten signature verification systems adapted to mobile scenarios, i.e., using mobile devices such as smartphones and tablets with the finger as input, achieving EERs around 20.0\%. In~ \cite{2016_IEEE_THMS_DoodlePass_Marcos}, the authors proposed the use of graphical doodles and pseudosignatures (i.e., simplified versions of the signatures drawn with the finger). EERs above 20.0\% were obtained in both cases for imitation attacks.

Finally, our proposed approach has been compared to other state-of-the-art authentication systems based on handwritten passwords. In~\cite{Travieso_2015}, the authors proposed the use of handwritten passwords with a fixed length of 8 characters, achieving a final False Acceptance Rate (FAR) of 10.42\% when using  a total of 12 training samples per user (the False Rejection Rate FRR was not provided by the authors). In~\cite{Sae_Bae_2017}, Nguyen \textit{et al.} evaluated the potential of drawing each digit of a 4-digit PIN one by one, achieving a final result of 4.84\% EER when considering a total of 5 enrolment samples. Our proposed approach achieves a final value of 3.8\% EER and it is able to mitigate the limitations of~\cite{Travieso_2015} about the size of the touchscreen, as users perform numerical digits one at a time. Additionally, we only consider 3 enrolment samples and not 5 as in~\cite{Sae_Bae_2017} in order to improve the usability of our approach.

\section{Password Generation and System Setup}\label{password_real_systems}
In this section we discuss specific details for the deployment of our proposed approach in real scenarios considering the same experimental protocol described in Sec. \ref{experimental_protocol}. The DTW Adapted System has been considered for this analysis.

First, we focus on PIN-based systems. For this scenario, we propose to use passwords based on 4 digits as users have to memorize them and it is not feasible from the point of view of the user to consider longer passwords. Regarding the enrolment stage, we propose to request 3 enrolment samples per digit to each user. We consider this as something feasible for real applications as users would have to perform a total of 4 digits $\times$ 3 samples/digit $=$ 12 samples, i.e., 12 samples $\times$ 2 seconds/sample $\simeq$ 25 seconds. 

\begin{figure}[tb]
  \centering
    \includegraphics[width=0.9\linewidth]{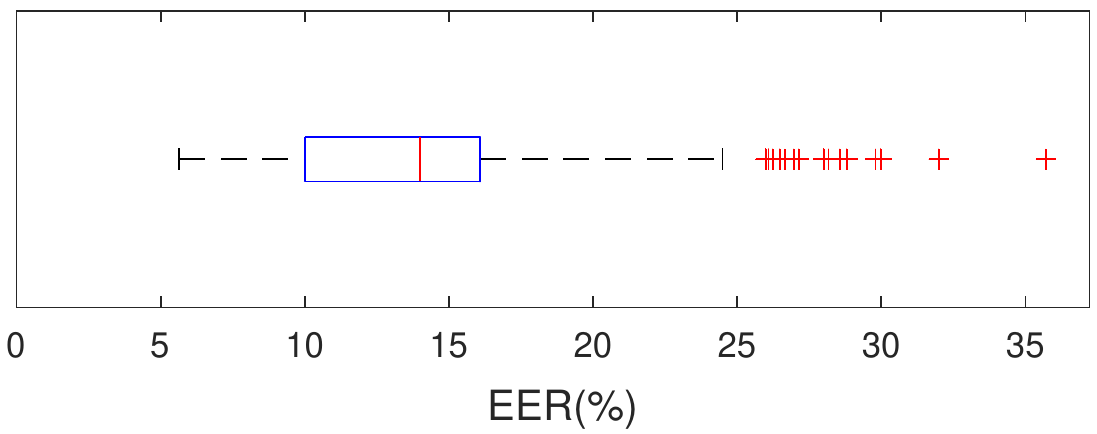}
  \caption{\textbf{PIN System}: Boxplot for the case of considering all 4-digit password combinations. On the box, the central mark indicates the median, and the left and right edges of the box indicate the 25th and 75th percentiles, respectively.}
  \label{PIN_boxplot}
\end{figure}

Once we have fixed the number of enrolment samples and digits parameters, we design what type of passwords we let users to use (i.e., we design the Password Generation module in Fig.~\ref{diagram_main}). The following cases are considered regarding both the system performance and number of possible combinations: \textit{i)} ALL password combinations are allowed, and \textit{ii)} only combinations using the BEST 4 digits selected in Table~\ref{exp3_enrolment_length} and with no repetitions (recall in Sect.~\ref{longitud_enrolment_analysis} we obtained that the most discriminative password combinations in terms of touch biometric information didn't include repeated digits). Fig.~\ref{PIN_boxplot} shows the EER distribution values obtained for all possible password combinations. On the box, the central mark indicates the median, and the left and right edges of the box indicate the 25th and 75th percentiles, respectively. The whiskers extend to the most extreme data points not considered outliers, and the outliers are plotted individually. In general, we can see that the 75\% of password combinations provide results below 16.2\% EER. Analyzing the case ALL, the system performance results achieved are between 5.9\% and 35.7\% EER with a total of $10^4$ combinations. The performance is improved in the case BEST with a 5.9\% EER for all considered combinations. However, users would be able to choose only among $4!$ combinations (i.e., 24). Besides, the security level of the first authentication stage would decrease as fewer password combinations would be possible. Therefore, a good choice could be to select all possible passwords that provide results in a range of EERs. For example, permitting between 5.9\% and 10.0\% EER. This approach would allow users to choose among 2,956 different 4-digit passwords.


\begin{table}[tb]
\centering
\caption{\textbf{OTP System}: number of 7-digit possible combinations and system performance results.}
\label{OTP_system_analysis}
\begin{adjustbox}{width=0.45\textwidth}
\begin{tabular}{c|c|c|}
\cline{2-3}
                                    & \# Password Combinations & EER(\%)    \\ \hline
\multicolumn{1}{|c|}{Case ALL}  & $10^7$    & 3.8 to 14.0 \\ \hline
\multicolumn{1}{|c|}{Case BEST} & 5,040                       & 3.8        \\ \hline
\end{tabular}
\end{adjustbox}
\end{table}

Now, we analyze the OTP-based system. For this scenario, we propose to use passwords composed of 7 digits, similar to current OTP-based applications, as users do not have to memorize the password, i.e., the system is in charge of selecting and providing different passwords to the user each time is required. Regarding the enrolment stage, we also propose to request 3 enrolment samples per digit so users would have to perform a total of 10 digits $\times$ 3 samples/digit $=$ 30 samples, i.e., 30 samples $\times$ 2 seconds/sample $\simeq$ 1 minute.

Once we have fixed both the number of enrolment samples and the length of the password, we analyze the content of the passwords. For this scenario, the following cases are considered: \textit{i)} ALL digit combinations are allowed, and \textit{ii)} only combinations using the BEST 7 digits selected in Table~\ref{exp3_enrolment_length} with no repetitions. Table~\ref{OTP_system_analysis} depicts the number of possible combinations as well as the EER (\%) for both cases. Analyzing the case in which users can choose any possible combination, the system performance results achieved are between 3.8\% and 14.0\% EER. However, it is important to remark that for this case (longer passwords) results were obtained due to experimental restrictions using the SFFS algorithm and limiting the maximum number of digit repetitions to 4, so the final 14.0\% EER might get a bit worse in practice when considering all possible digit combinations. This approach is further improved in the case BEST with a final 3.8\% EER. For this scenario we propose to use this second case as there would be a total of $7!$ (i.e., 5,040) combinations that provide the best system performance for our proposed touch biometric approach.

\section{Conclusions}\label{conclusions}
This work evaluates the advantages and potential of incorporating handwritten touch biometrics to password-based mobile authentication systems. The new e-BioDigit database that comprises handwritten numerical digits from 0 to 9 is used in the experiments reported in this work and it is available together with benchmark results in GitHub\footnote{https://github.com/BiDAlab/eBioDigitDB}. Data were collected in two sessions for a total of 93 subjects. Handwritten numerical digits were acquired using the finger touch as the input on a Samsung Galaxy Note 10.1 general purpose tablet device. 

For the new e-BioDigit database, we report a benchmark evaluation using two different state-of-the-art approaches: \textit{i)} DTW in combination with the SFFS function selection algorithm, and \textit{ii)} RNN deep learning technology. Both approaches have been compared, achieving very good results even for the case of using just a single enrolment sample. In addition, we perform a complete analysis of the touch biometric system regarding the discriminative power of each handwritten digit, and the robustness of our proposed approach when increasing the length of the password and the number of enrolment samples per user.

Our proposed approach achieves good results with EERs ca. 4.0\% when considering imitation attacks, outperforming other traditional biometric verification traits such as the handwritten signature or graphical passwords on similar mobile scenarios. Additionally, we discuss specific details for the deployment or our proposed approach on current PIN- and OTP-based authentication systems.

Future work will be oriented to enlarge the current e-BioDigit database in order to consider lower- and upper-case letters and also to train more complex deep learning architectures.

\section*{Acknowledgments}
This work has been supported by projects: BIBECA (MINECO), Bio-Guard (Ayudas Fundaci\'on BBVA a Equipos de Investigaci\'on Cient\'ifica 2017) and by UAM-CecaBank. Ruben Tolosana is supported by a FPU Fellowship from Spanish MECD.



%

%

{
\bibliographystyle{IEEEtran}
\bibliography{egbib2}
}

\begin{IEEEbiography}[{\includegraphics[width=1in,height=1.25in,clip,keepaspectratio]{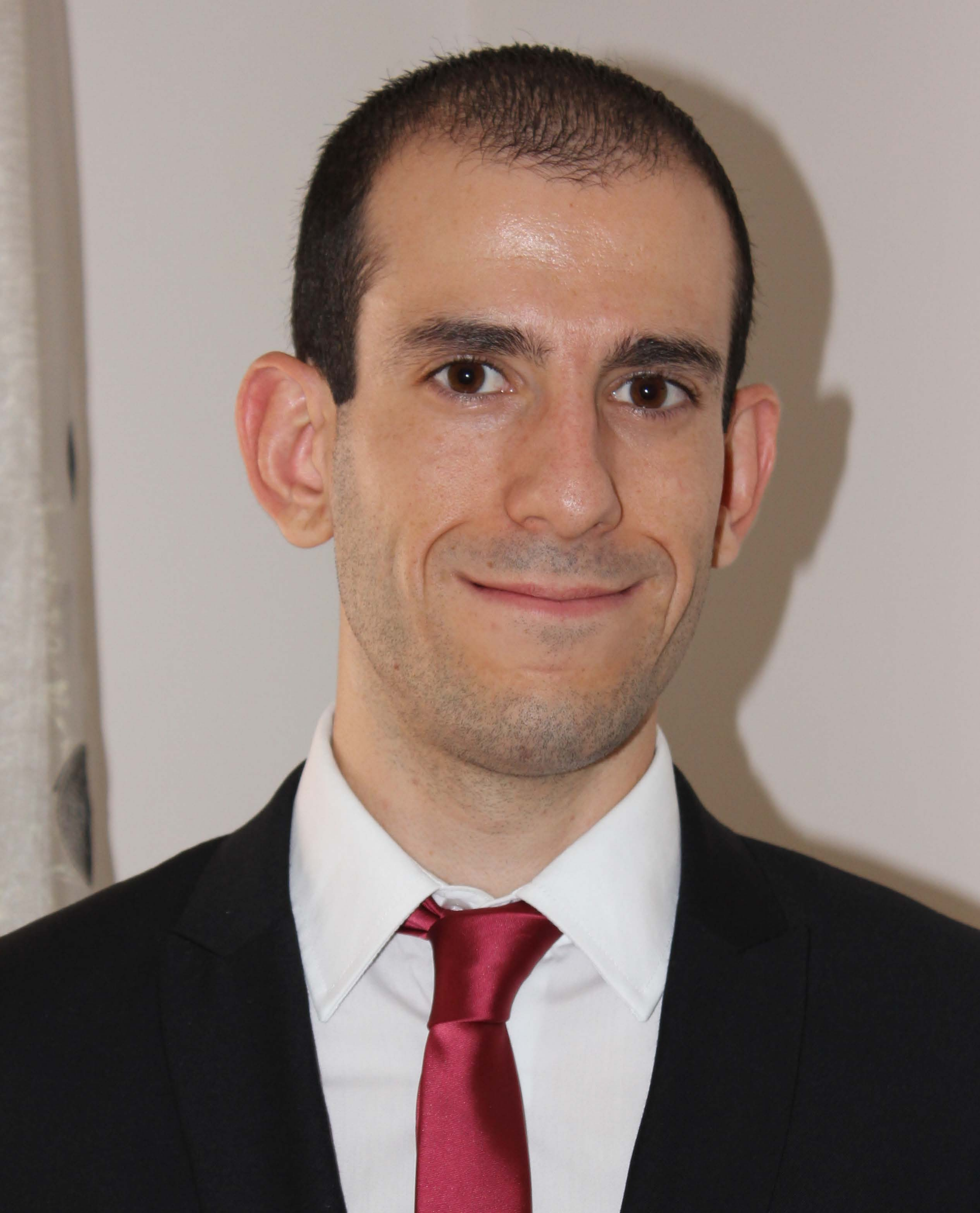}}]{Ruben Tolosana} received the M.Sc. degree in Telecommunication Engineering in 2014 from Universidad Autonoma de Madrid. In April 2014, he joined the Biometrics and Data Pattern Analytics - BiDA Lab at the Universidad Autonoma de Madrid, where he is currently collaborating as an assistant researcher pursuing the Ph.D. degree. Since then, Ruben has been granted with several awards such as the FPU research fellowship from Spanish MECD (2015), and the European Biometrics Industry Award from EAB (2018). His research interests are mainly focused on signal and image processing, pattern recognition, deep learning, and biometrics, particularly in the areas of handwriting and handwritten signature. He is author of several publications and also collaborates as a reviewer in many different international conferences (e.g., ICDAR, ICB, EUSIPCO, etc) and high-impact journals (e.g., IEEE Transactions of Information Forensics and Security, IEEE Transactions on Cybernetics, ACM Computing Surveys, etc). Finally, he has participated in several National and European projects focused on the deployment of biometric security through the world.
\end{IEEEbiography}

\begin{IEEEbiography}[{\includegraphics[width=1in,height=1.25in,clip,keepaspectratio]{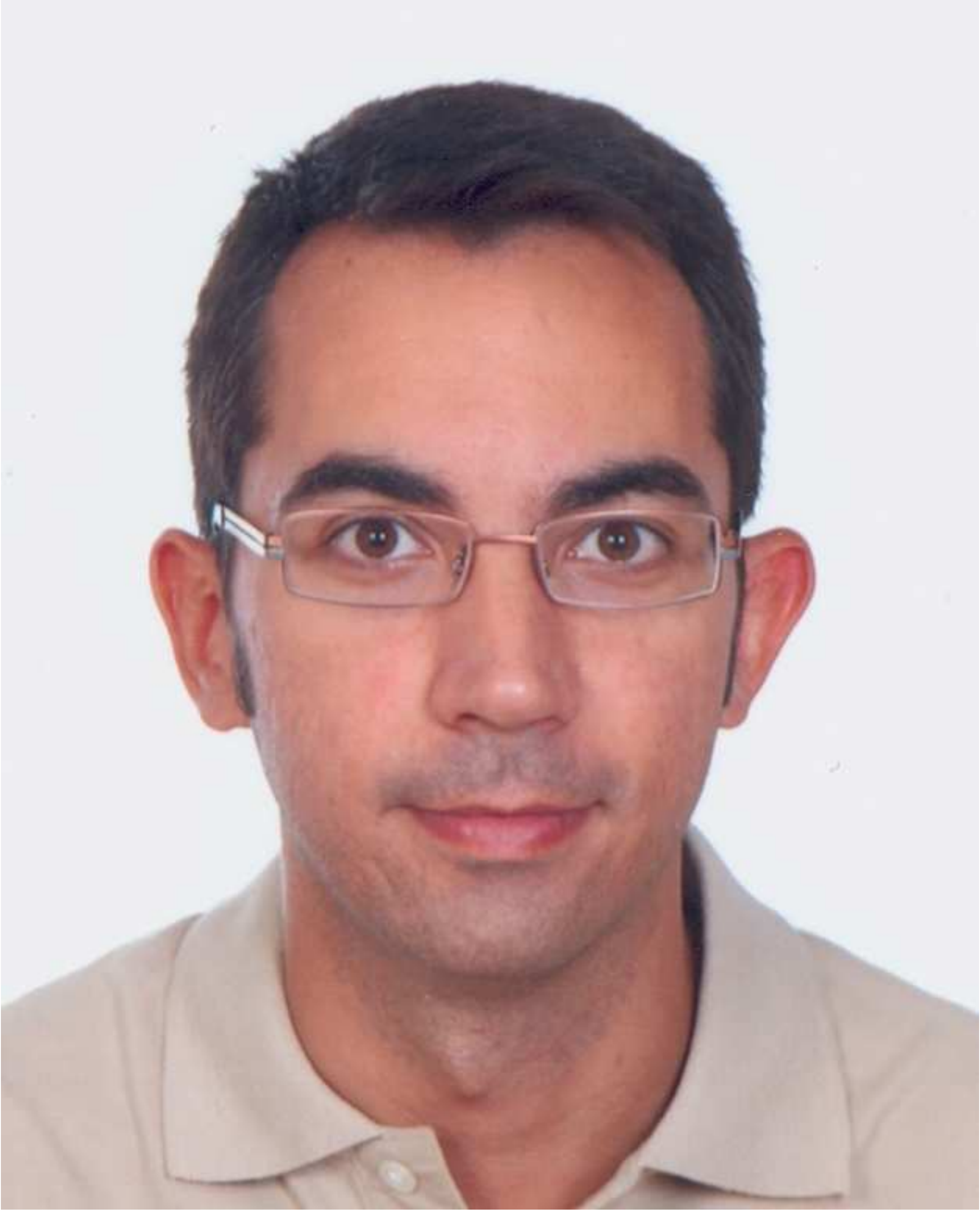}}]{Ruben Vera-Rodriguez} received the M.Sc. degree in telecommunications engineering from Universidad de Sevilla, Spain, in 2006, and the Ph.D. degree in electrical and electronic engineering from Swansea University, U.K., in 2010. Since 2010, he has been affiliated with the Biometric Recognition Group, Universidad Autonoma de Madrid, Spain, where he is currently an Associate Professor since 2018. His research interests include signal and image processing, pattern recognition, and biometrics, with emphasis on signature, face, gait verification and forensic applications of biometrics. He is actively involved in several National and European projects focused on biometrics. Ruben has been Program Chair for the IEEE 51st International Carnahan Conference on Security and Technology (ICCST) in 2017; and the 23rd Iberoamerican Congress on Pattern Recognition (CIARP 2018) in 2018.

\end{IEEEbiography}

\begin{IEEEbiography}[{\includegraphics[width=1in,height=1.25in,clip,keepaspectratio]{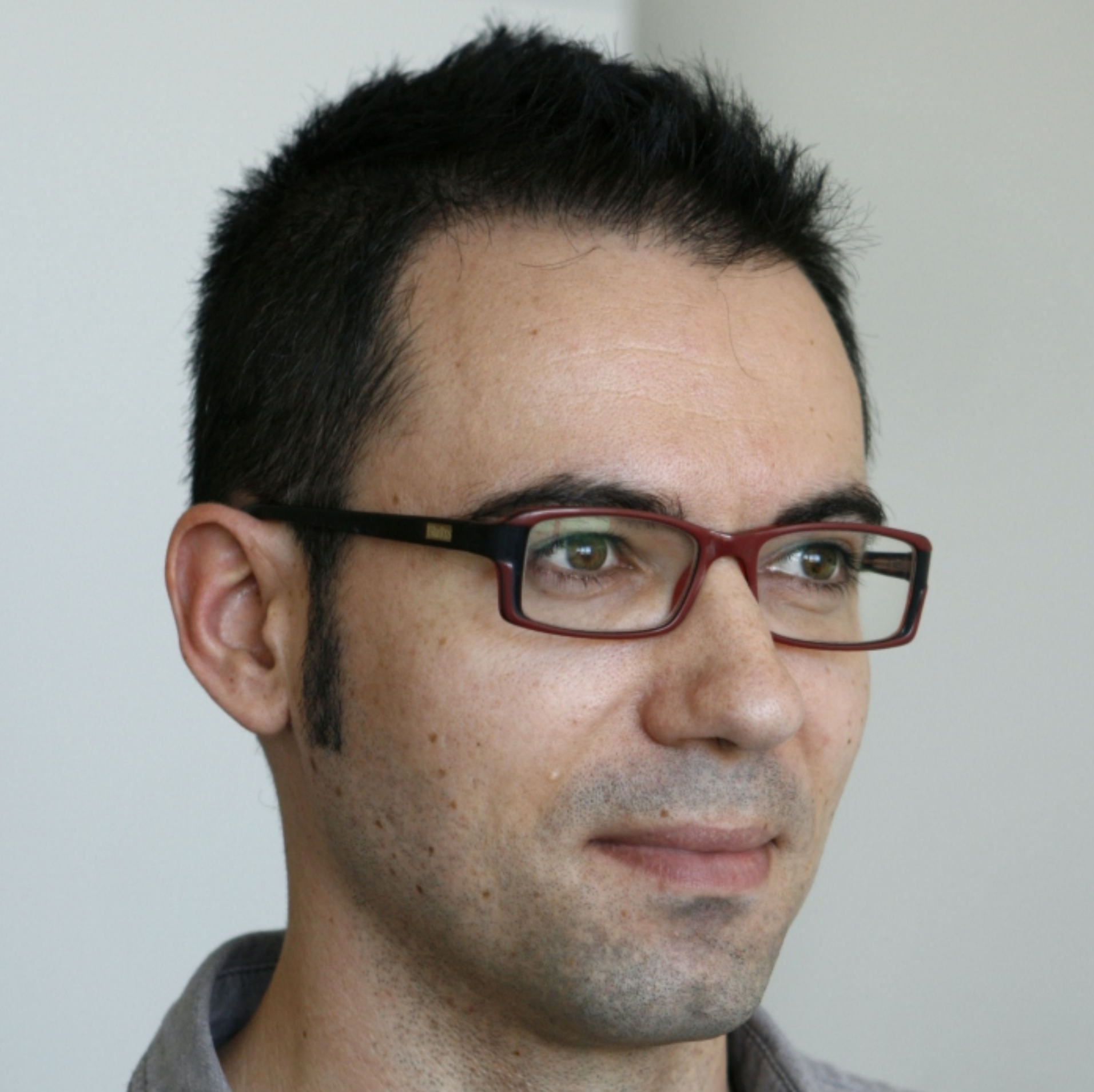}}]{Julian Fierrez} received the M.Sc. and Ph.D. degrees in telecommunications engineering from the Universidad Politecnica de Madrid, Spain, in 2001 and 2006, respectively. Since 2002, he has been with the Biometric Recognition Group, Universidad Politecnica de Madrid. Since 2004, he has been with the Universidad Autonoma de Madrid, where he is currently an Associate Professor. From 2007 to 2009, he was a Visiting Researcher with Michigan State University, USA, under a Marie Curie Fellowship. His research interests include signal and image processing, pattern recognition, and biometrics, with an emphasis on multibiometrics, biometric evaluation, system security, forensics, and mobile applications of biometrics. He has been actively involved in multiple EU projects focused on biometrics (e.g., TABULA RASA and BEAT), and has attracted notable impact for his research. He was a recipient of a number of distinctions, including the EAB European Biometric Industry Award 2006, the EURASIP Best Ph.D. Award 2012, the Miguel Catalan Award to the Best Researcher under 40 in the Community of Madrid in the general area of science and technology, and the 2017 IAPR Young Biometrics Investigator Award. He is an Associate Editor of the IEEE TRANSACTIONS ON INFORMATION FORENSICS AND SECURITY and the IEEE TRANSACTIONS ON IMAGE PROCESSING.
\end{IEEEbiography}

\end{document}